\documentclass[a4paper, onecolumn, unpublished]{quantumarticle}

\pdfoutput=1

\usepackage[utf8]{inputenc}
\usepackage[english]{babel}
\usepackage[T1]{fontenc}
\usepackage{amsmath}
\usepackage{amssymb}
\usepackage{bbold}
\usepackage{hyperref}
\usepackage{stmaryrd}
\usepackage{algpseudocode}
\usepackage{algorithm}
\usepackage{xcolor}
\usepackage{qcircuit}
\usepackage[numbers,sort&compress]{natbib}

\definecolor{mR}{RGB}{136,34,85}
\definecolor{mG}{RGB}{13,89,38}
\definecolor{mB}{RGB}{51,34,136}

\newcommand{\id}{\mathbb{1}}
\newcommand\Tstrut{\rule{0pt}{3ex}}
\newcommand{\returnarrow}{\rotatebox[origin=c]{180}{\(\Lsh\)}}

\title{Planar Floquet Codes}
\author{Christophe Vuillot}
\email{christopĥe.vuillot@inria.fr}
\affiliation{Inria Nancy}
\begin{document}
	\maketitle
	
	\abstract{
		A protocol called the ``honeycomb code'', or generically a ``Floquet code'', was introduced by Hastings and Haah in \cite{hastings_dynamically_2021}.
		The honeycomb code is a subsystem code based on the honeycomb lattice with zero logical qubits but such that there exists a schedule for measuring two-body gauge checks leaving enough room at all times for two protected logical qubits.

		In this work we show a way to introduce boundaries to the system which curiously presents a rotating dynamics but has constant distance and {\color{red}is therefore not fault-tolerant}.

	\section{Introduction}
	Building fault-tolerant quantum computers is an extremely challenging task.
	Bridging the gap between theoretical fault-tolerant schemes and physics experiment is an ongoing global effort.
	In this effort the honeycomb code was introduced in \cite{hastings_dynamically_2021}.
	It is a protocol for a fault tolerant quantum memory which uses only two-qubit Pauli measurements.
	According to numerical simulations \cite{gidney_fault-tolerant_2021}, it seems it could be a promising contender to the long standing favorite, the surface code \cite{kitaev_fault-tolerant_2003, dennis_topological_2002}, if one has access to native two-qubit measurements.
	The question of designing boundaries and a planar geometry for the honeycomb code was left open in \cite{hastings_dynamically_2021, gidney_fault-tolerant_2021}.
	In this paper we present a way of adding boundaries which generate a curious rotating dynamics for the logical operators but which is not fault-tolerant as there are constant size space-time logical operators.
	In a subsequent paper \cite{haah_boundaries_2021} a fault-tolerant way to introduce boundaries is presented.
	
	In Section~\ref{sec:floquetfromcolorcodes} we show how to define general Floquet codes and a way to introduce boundaries to them.
	Then in Section~\ref{sec:planarfloquetcodes} we focus on specific instances of planar Floquet codes.
	
	\section{Floquet Codes from 2D Color Codes}
	\label{sec:floquetfromcolorcodes}
	In \cite{hastings_dynamically_2021} the authors define Floquet codes using the hexagonal lattice wrapped around a torus which takes inspiration from Kitaev's honeycomb model \cite{kitaev_anyons_2006}.
	In the same way that Kitaev's honeycomb model can be defined on more general geometries \cite{suchara_constructions_2011}, Floquet codes can be defined on any 2D color code lattice \cite{bombin_topological_2006}.
	We will use a slightly different definition compared to \cite{hastings_dynamically_2021} which is equivalent up to a Clifford unitary.
	
	We show in Section~\ref{sec:defs}~and~\ref{sec:schedule} that a planar graph \(\mathcal{G}=(V, E, F)\) tiling a surface with no boundaries, which is 3-valent and whose faces are 3-colorable defines a \(\llbracket n, k, d\rrbracket\) Floquet code.
	The parameters are given by \(n=\vert V\vert\), \(k\) is obtained from the genus of the surface as in Eq.~\eqref{eq:logdim} and \(d\) is a fraction of the minimal length of a homologically non-trivial cycle in the graph.
	The code space is stabilized by measuring gauge checks corresponding to edges following a cyclic schedule defined by the three colors.
	We then show in Section~\ref{sec:bound} how to properly introduce boundaries to the system.
	
	\subsection{Gauge Checks, Stabilizers and Inner Logical Operators}
	\label{sec:defs}
	
	\begin{figure}[ht!]
		\centering
		\includegraphics[width=.4\linewidth]{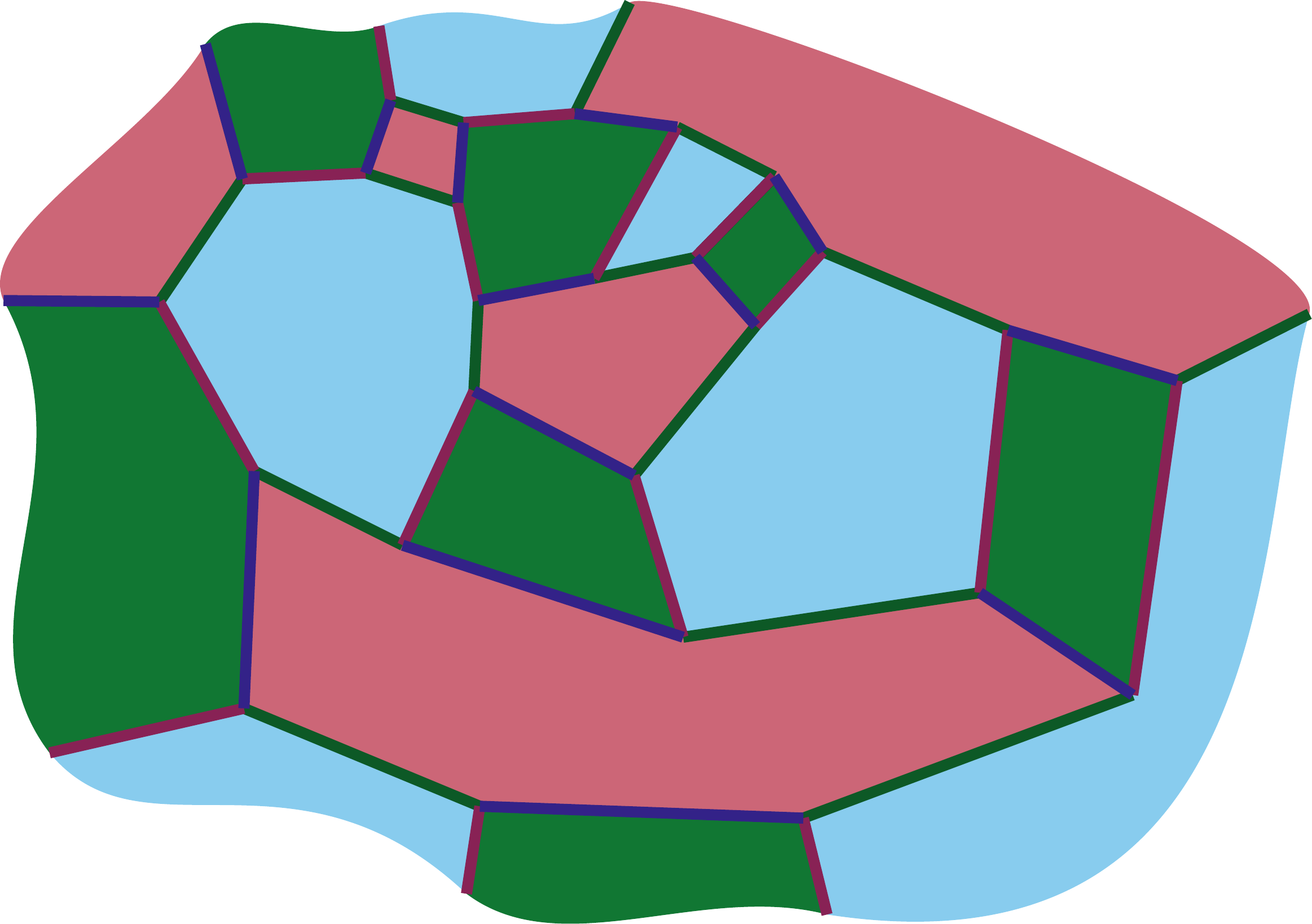}
		\caption{Part of the bulk of a generic planar graph suitable to define a Floquet code.}
		\label{fig:generic}
	\end{figure}
	
	We start from a planar graph, \(\mathcal{G} = (V, E, F)\), with vertex set \(V\), edge set \(E\subset V^2\) and face set \(F\subset2^V\).
	We impose that \(\mathcal{G}\) is tiling a 2D surface with the property required to define a color code: Namely vertices are 3-valent and the faces are 3-colorable.
	We also consider that \(\mathcal{G}\) as a single connected component.
	We denote $\vert F\vert = n_f$ the number of faces, $\vert V\vert = n_v$ the number of vertices and $\vert E\vert = n_e$ the number of edges.
	For now we consider that the surface does not have boundaries but we will introduce some later.
	We use red, green and blue (\(R, G, B\)) for the three colors and a coloring map for the faces, \(C : F \rightarrow \{R, G, B\}\).
	The faces are three colorable which is expressed by
	\begin{equation}
		\forall (f_1, f_2)\in F^2,\, f_1\cap f_2\in E \Rightarrow C(f_1)\neq C(f_2).
	\end{equation}
	The coloring map can be extended to edges as follows
	\begin{align}
		C : E &\rightarrow \{R, G, B\}\nonumber\\
			  e &\mapsto C(e)\in \{R, G, B\} \setminus \{C(F(e))\},
	\end{align}
	where \(F(e)\) designate the faces containing \(e\) (the coboundary of \(e\)).
	In words, an edge is colored with the complementary color to the two colors of its bordering faces.
	Said differently, edges are of the same color as the faces they link, see an example drawn in Figure~\ref{fig:generic}.

	\subsubsection{Gauge Checks}
	Qubits are put on vertices and to each color we associate a Pauli label.
	Throughout we use the choice 
	\begin{equation}
		P^c = \left \{\begin{matrix}
			X,&\text{ for } c=R\\
			Y,&\text{ for } c=G\\
			Z,&\text{ for } c=B\\
		\end{matrix}\right .
	\end{equation}
	For \(v\in V\) and \(P\in\{X, Y, Z\}\), we write \(P_v\) for the Pauli operator acting as \(P\) on the qubit associated with \(v\) and the identity on the other qubits.
	Each edge \(e = (v_1,v_2)\) defines a gauge check, \(P_e\) using the Pauli label associated with its color
	\begin{equation}
		P_e = P^{C(e)}_{v_1}P^{C(e)}_{v_2}.
	\end{equation}
	That is to say red edges are $X$-checks, green edges $Y$-checks and blue edges $Z$-checks.
	We have one linear dependency between the gauge checks:
	\begin{equation}
		\prod_{e\in E}P_e = \prod_{v\in V}X_vY_vZ_v \propto \id.
	\end{equation}
	There is exactly one such dependency per connected component as for each qubit to be acted on by the identity the only way is to not have any of its edges or the three.
	We restricted ourselves to graphs with exactly one connected component so we count one linear dependency.
	
	Using these gauge checks we define a basis of Pauli operators for the Hilbert space of \(n_v\) qubits using the notion of subsystem code.
	The gauge checks form the gauge group whose dimension, \(n_g\), is given by
	\begin{equation}
		n_g = n_e - 1.\label{eq:gaugedim}
	\end{equation}

	\subsubsection{Stabilizers}
	The center of the gauge group contains all operators generated by gauge checks commuting with every gauge check.
	For a product of gauge checks to commute with all gauge checks it is necessary and sufficient that the product runs over cycles of edges in the graph.
	Homologically trivial cycles are generated by faces of the graph.
	We call these stabilizers.
	A face \(f\in F\) of color \(C(f)\) defines a stabilizer \(P_f\) as follows
	\begin{equation}
		P_f = \prod_{v\in f}P^{C(f)}_v \propto \prod_{e\in f}P_e .
	\end{equation}
	That is to say that red faces are \(X\)-stabilizers, green faces are \(Y\)-stabilizers and blue faces are \(Z\)-stabilizers.
	Since there are no boundaries and the graph has one connected component, there is one linear dependency between stabilizers as the product of all faces is proportional to the identity
	\begin{equation}
		\prod_{f\in F} P_f = \prod_{v\in V}X_vY_vZ_v \propto \id.
	\end{equation}
	
	\subsubsection{Inner Logical Operators}
	Homologically non-trivial cycles are also in the center of the gauge group.
	In \cite{hastings_dynamically_2021}, they are called inner logical operators and their number depend on the orientability and genus of the surface.
	We denote as \(k\) their number which is given by:
	\begin{equation}
		k = \left \{\begin{matrix}
			2g,&\text{ if orientable surface of genus $g$}\\
			g,&\text{ if non-orientable surface of (non-orientable) genus g}
		\end{matrix}\right ..\label{eq:logdim}
	\end{equation}
	
	In summary the dimension of the center of the gauge group, \(n_s\), is given by
	\begin{equation}
		n_s = n_f - 1 +k.\label{eq:stabdim}
	\end{equation}
	
	\subsubsection{Empty Subsystem Code}
	We can show that these operators completely stabilize the space.
	That is to say that there are zero logical qubits if we consider inner logical operators as stabilizers.
	Note that this fact has already been detailed and observed in \cite{suchara_constructions_2011}.
	
	This is computed as follows.
	Since the vertices are 3-valent, we have that
	\begin{equation}
		3n_v = 2n_e.\label{eq:trivalent}
	\end{equation}
	Euler characteristics gives us
	\begin{equation}
		n_v - n_e + n_f = 2 - k.\label{eq:euler}
	\end{equation}
Using Eq.~\eqref{eq:trivalent} in Eq.~\eqref{eq:euler} we deduce a relation between the number of edges and the number of faces, namely
\begin{equation}
	n_e = 3n_f + 3k -6.\label{eq:nenp}
\end{equation}

	The number of logical qubits, \(n_L\), is then given by
	\begin{align}
		n_L &= n_v - \frac{n_g - n_s}{2} - n_s\nonumber\\
			   &= n_e - n_f + 2 - k - \frac{n_e - 1 + n_f - 1 + k}{2}\nonumber\\
			   &= \frac{n_e - 3n_f - 3k + 6}{2}\nonumber\\
		n_L &= 0.
	\end{align}
	Hence there are zero logical qubits in this subsystem code.
	
	\subsection{Measurement Schedule and Instantaneous Stabilizer Group}
	\label{sec:schedule}
	As explained in \cite{hastings_dynamically_2021}, in order to use these gauge checks to protect quantum information it is possible to measure them in a specific order such that one effectively measures the face stabilizers but never the inner logical operators (homologically non-trivial cycles).
	This will ensure that at all times there is room for \(k\) logical qubits in the stabilized subspace.

	The idea consists in measuring all checks of a given color at once at each step and cycle through the 3 colors giving a cyclic 3-steps schedule.
	When measuring say red (\(X\)) checks after having measured green (\(Y\)) checks we learn the value of each cycle composed of red and green checks.
	We call these type of cycles bi-colored cycles.
	Bi-colored cycles involve an even number of qubits and a Pauli operator acting like \(Z\) on every qubit of such cycle can therefore be measured by two successive commuting measurements of type \(Y\) then \(X\) \cite{suchara_constructions_2011}.
	
		\begin{figure}[ht!]
		\centering\includegraphics[width=.3\linewidth]{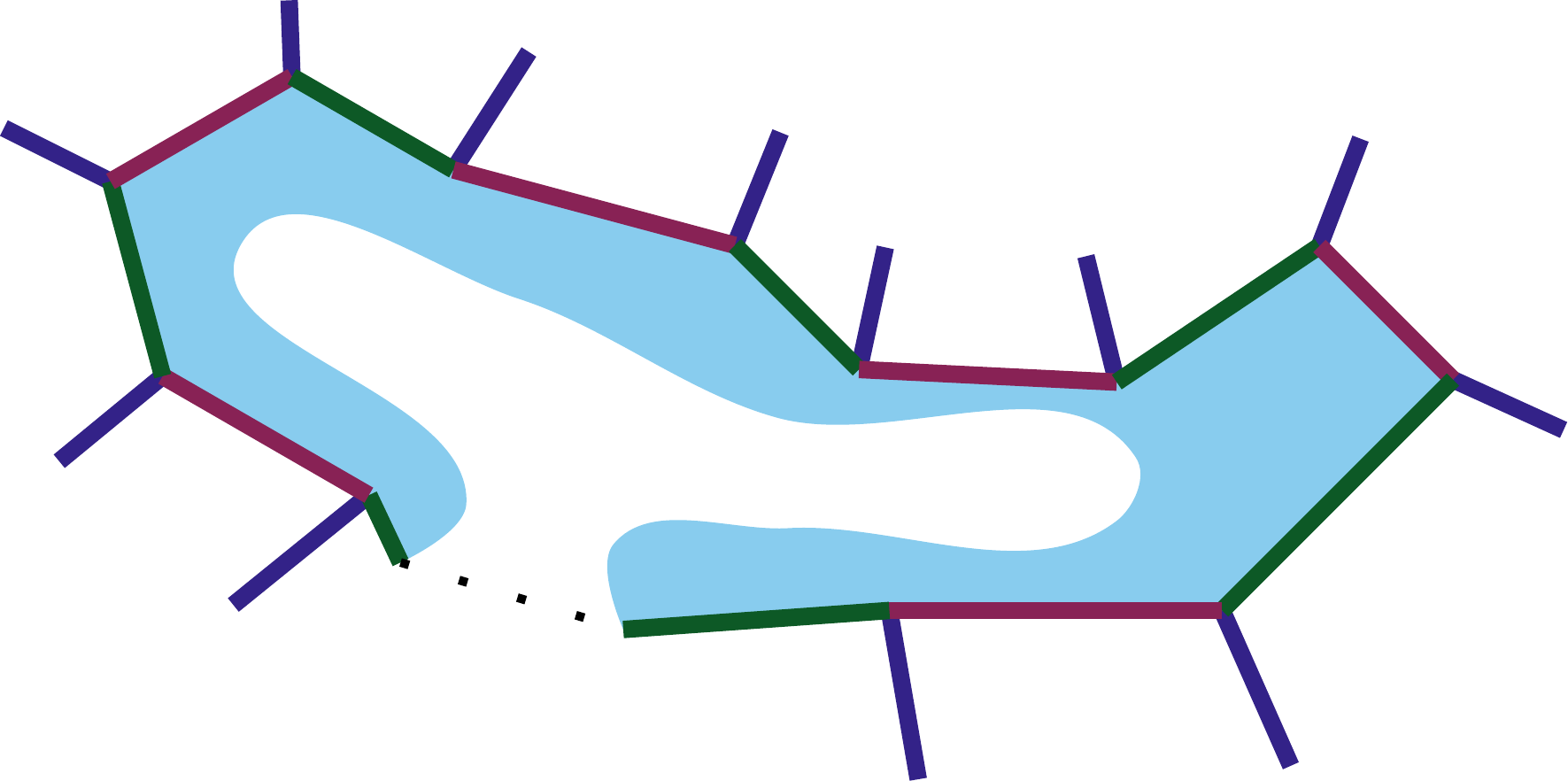}
		\caption{Illustration of the fact that bicolored cycles are necessarilly trivial when there are no boundaries and that no face borders a non-trivial loop.}
		\label{fig:trivialbicolored}
	\end{figure}
We can show that these bi-colored cycles are necessarily homologically trivial when there are no homologically non-trivial faces and therefore cannot be inner logical operators.
	Indeed, pick a cycle of green and red edges.
	Take a starting green edge followed by a red edge of the cycle.
	they intersect at a qubit which has a third blue edge going out of it.
	This blue edge points at a blue face bordering both green and red edges we started with.
	The other end of the red edge has a blue and a green edge adjacent.
	The blue one cannot belong to the blue face and therefore the next green edge in the cycle has to belong to the same blue face.
	following the whole cycle this way we see that this cycle borders a single blue face.
	Since there are no boundaries, and if we assume that there are no homologically non-trivial faces, this cycle is trivial.
	The reasoning is illustrated in Figure~\ref{fig:trivialbicolored}.
	Note that this also implies that inner logical operators involve edges of all three colors.

	In conclusion, the natural measurement schedule allows to measure the stabilizers but never the inner logical operators.
	We sumarize the schedule as well as the instantaneous stabilizer group (ISG) at each step in Algorithm~\ref{alg:schedule}.
	The set of red gauge checks is denoted \(R\)-gauge, the set of red stabilizers is denoted \(R\)-stab. and similarly for blue with \(B\) and green with \(G\).

	Similarly to what is shown in \cite{wootton_family_2015, hastings_dynamically_2021} in the steady state regime and at each step the instantaneous stabilizer group (ISG) is equivalent through a constant depth Clifford circuit to a 2D homological code.
	Indeed for each measured edge the two corresponding qubits are projected onto a one qubit subspace.
	For instance if we just measured all red \(X\)-checks, each red edge is projected onto an effective qubit characterized by the Pauli operators \(X\otimes\id\) (equivalently \(\id\otimes X\)) and \(Y\otimes Y\) (equivalently \(Z\otimes Z\)).
	The face stabilizers can be seen as acting directly onto these effective qubits.
	The stabilizer group in this picture is that of a 2D homological code.
	This is detailed in Table~\ref{tab:homologicaleq} and illustrated in Figure~\ref{fig:genericsteps}.
		\begin{table}[ht!]
		\centering
		\begin{tabular}{c c |c c}
			\multicolumn{2}{c|}{Floquet code at step \(c\in\{R,G,B\}\)} & \multicolumn{2}{c}{equivalent homological code}\\\hline\Tstrut
			gauge checks: \(e\in E,\,C(e)=c\) & \(P_e = P^{c}\otimes P^{c} = \pm 1\) & edges (= qubits)&  \(\bar{Z} = P^c\otimes\id,\, \bar{X} = P^{\tilde{c}}\otimes P^{\tilde{c}}\)\\
			\(c\)-faces: \(f\in F,\, C(f) = c\) & \(P^c\)-stabilizers & faces & \(\bar{Z}\)-stabilizers\\
			other faces: \(f\in F,\, C(f) = \tilde{c}\neq c\) & \(P^{\tilde{c}}\)- stabilizers & vertices & \(\bar{X}\)-stabilizers
		\end{tabular}
		\caption{Correspondence of the stablized space after measuring gauge checks of color \(c\in\{R,G,B\}\) to an equivalent homological code.}
		\label{tab:homologicaleq}
	\end{table}
	\begin{figure}[ht!]
		\centering
		\includegraphics[width=.3\linewidth]{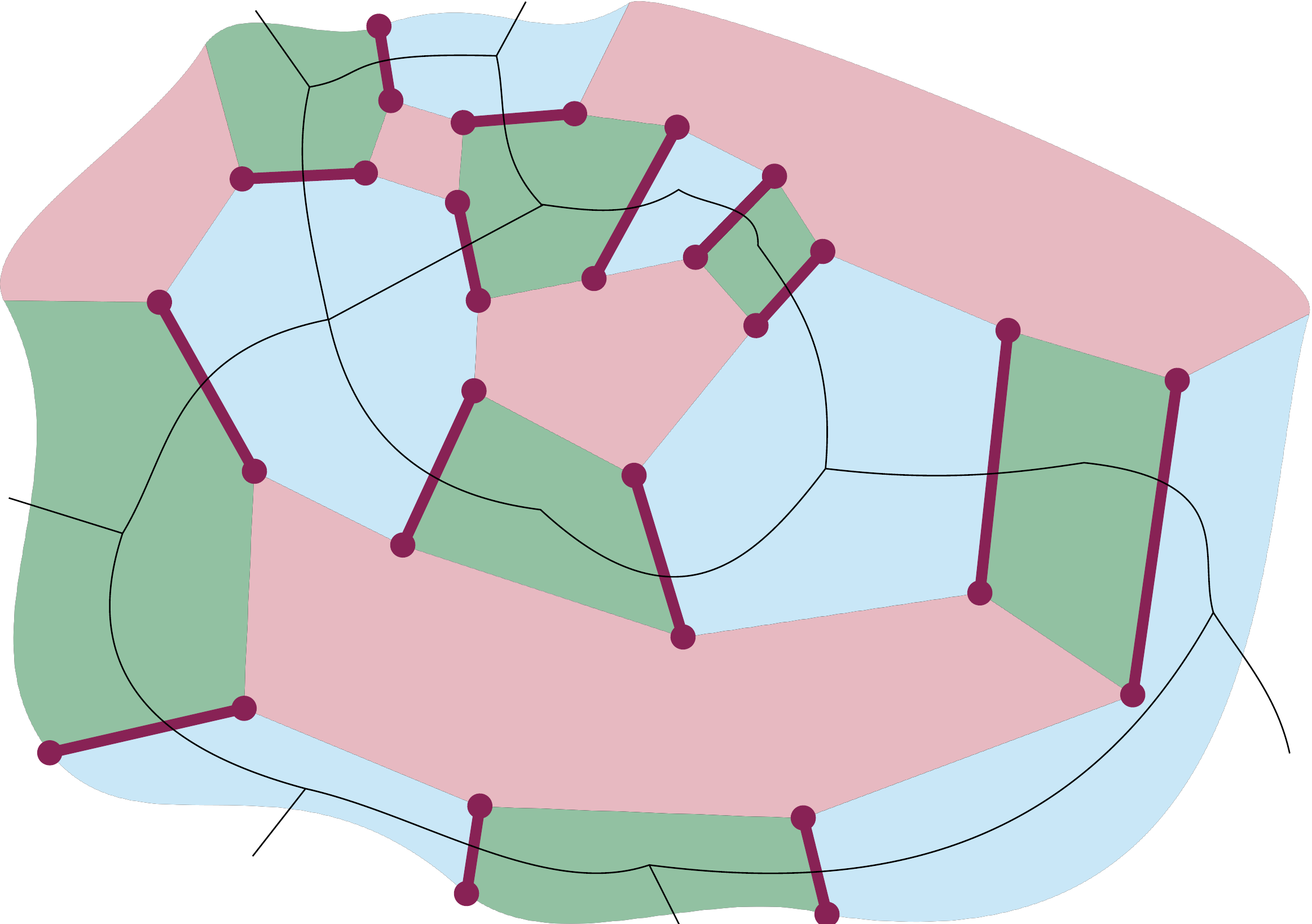}\hfil
		\includegraphics[width=.3\linewidth]{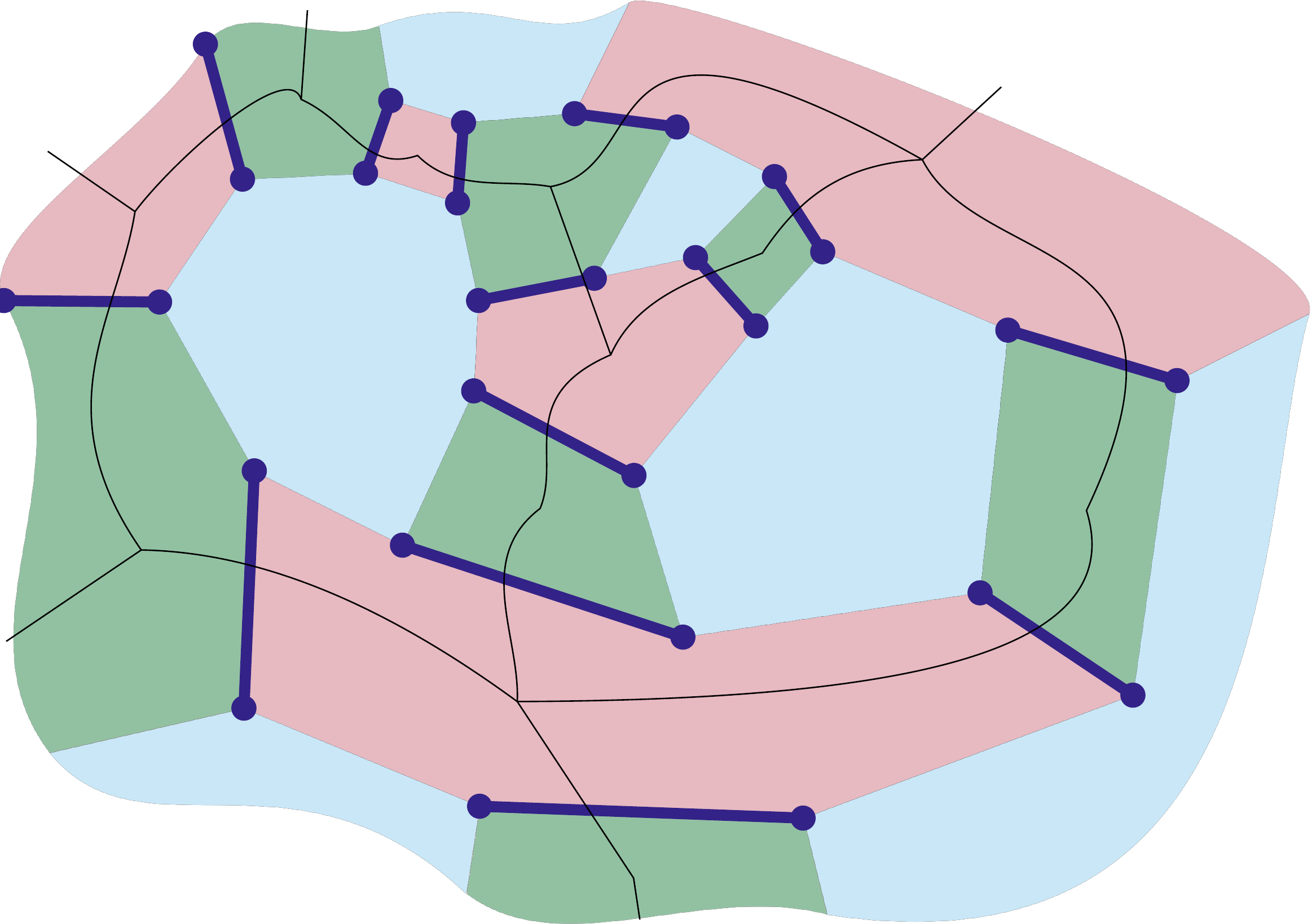}\hfil
		\includegraphics[width=.3\linewidth]{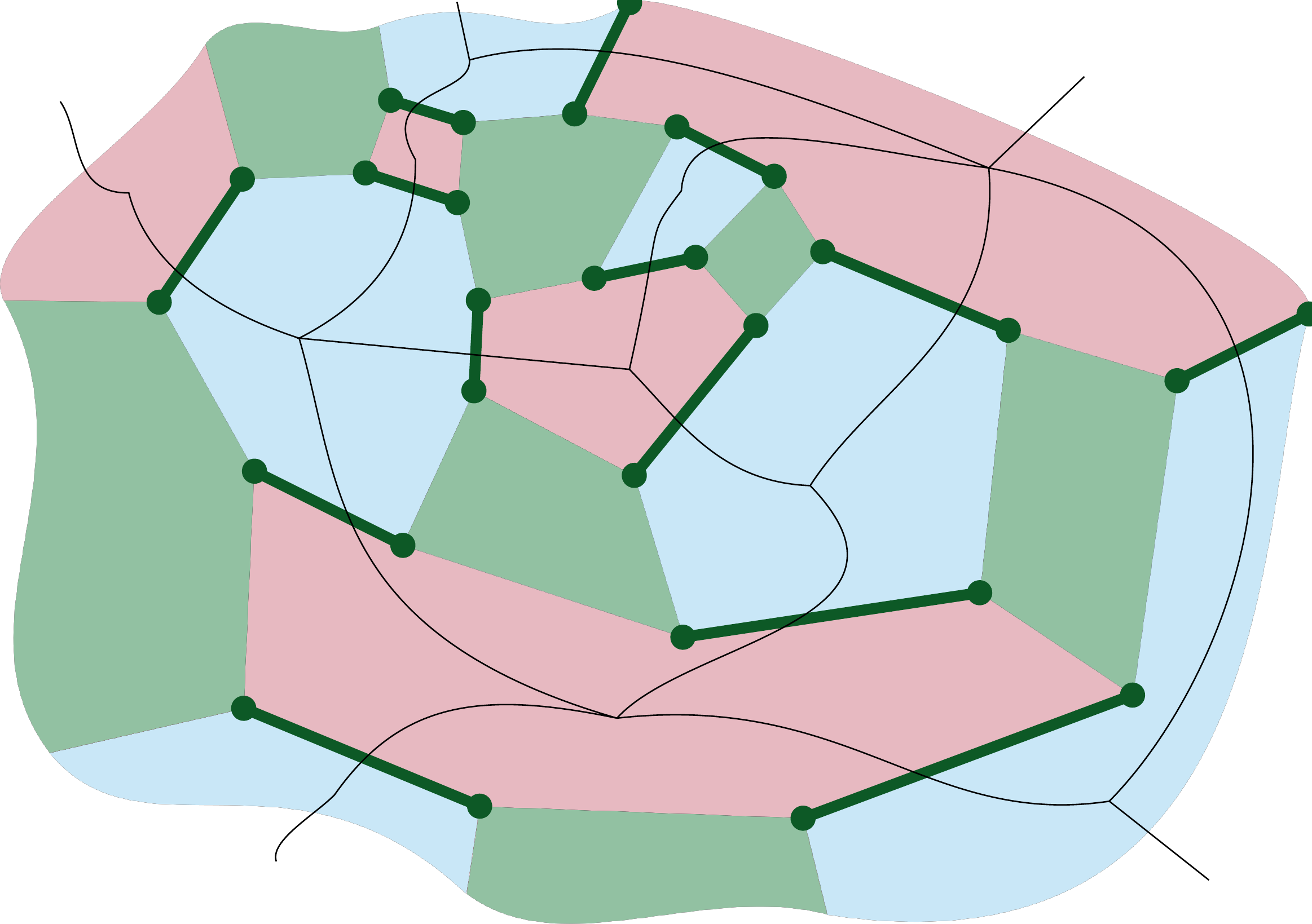}
		\caption{Representation of the three different configurations of stabilizers of the system at different time steps.
				The faces are stabilizers of the system at all times.
				The gauge checks that were just measured (and as such stabilize the space possibly with minus signs)  are represented by colored edges.
				The space stabilized is Clifford equivalent to a 2D homological code whose geometry is represented with thin black lines.
				In this homological code qubits are placed on the edges and stabilizers are defined by faces and vertices. }
			\label{fig:genericsteps}
	\end{figure}
	\begin{algorithm}[ht]
	\caption{Measurement Schedule}\label{alg:schedule}
	\begin{algorithmic}
		\State {\color{mR} \textbf{measure} red (\(X\)) gauge checks}\Comment{\(ISG = R\text{-gauge}\)}
		\State {\color{mB} \textbf{measure} blue (\(Z\)) gauge checks}
		\State {\color{mG}\returnarrow\textbf{record} green (\(Y\)) stabilizers}\Comment{\(ISG = B\text{-gauge}\cup G\text{-stab.}\)}
		\State {\color{mG} \textbf{measure} green (\(Y\)) gauge checks}
		\State {\color{mR} \returnarrow\textbf{record} red (\(X\)) stabilizers}\Comment{\(ISG = G\text{-gauge}\cup G\text{-stab.}\cup R\text{-stab.}\)}
		\Statex
		\Loop \Comment{Entering steady state}
		\State {\color{mR} \textbf{measure} red (\(X\)) gauge checks}
		\State {\color{mB} \returnarrow\textbf{record} blue (\(Z\)) stabilizers}\Comment{\(ISG = R\text{-gauge}\cup G\text{-stab.}\cup R\text{-stab.}\cup B\text{-stab.}\)}
		\State {\color{mB} \textbf{measure} blue (\(Z\)) gauge checks}
		\State {\color{mG} \returnarrow\textbf{record} green (\(Y\)) stabilizers}\Comment{\(ISG = B\text{-gauge}\cup G\text{-stab.}\cup R\text{-stab.}\cup B\text{-stab.}\)}
		\State {\color{mG} \textbf{measure} green (\(Y\)) gauge checks}
		\State {\color{mR} \returnarrow\textbf{record} red (\(X\)) stabilizers}\Comment{\(ISG = G\text{-gauge}\cup G\text{-stab.}\cup R\text{-stab.}\cup B\text{-stab.}\)}
		\EndLoop
	\end{algorithmic}
\end{algorithm}

	\subsection{Minimum Distance and Decoding}
	\label{sec:decoder}
	We have seen that homologically non-trivial cycles of gauge checks give logical operators for the code (the inner logical operators).
	Therefore the minimal length for them gives an upper-bound on the code distance.
	This set of logical operators does not constitute the full set of logical operators (they all commute with one another for instance).
	To form logical operators that are not generated by the gauge checks, one can use the equivalent homological code picture.
	One can construct logical operators there and translate them back to the Floquet code using the qubits definition in Table~\ref{tab:homologicaleq}.
	These logical operators are called outer logical operators in \cite{hastings_dynamically_2021}.
	Forming these, one can observe that the operators act on the edges of the original graph but can skip across faces of certain colors.
	For instance when the red checks are measured, a \(\bar{X}\)-logical, i.e. a co-cycle, also called electric operator, will act on a string of red edges that can skip through red faces.
	A \(\bar{Z}\)-logical operator, i.e. a cycle, also called magnetic operator, will act on blue and green edges but can skip across blue or green faces.
	These logical operators still extend over non-tricial cycles of the original edges but act only on a fraction of the vertices of the cycle.
	The worst fraction achievable depends on the size of the faces.
	If faces have bounded sizes then the minimum distance is at worst a constant fraction of the minimal length of a non-trivial cycle.
	
	A strategy for decoding the syndrome is given in \cite{hastings_dynamically_2021} and can readily be applied for general graphs.
	The main idea is to realize that Pauli errors happening between rounds can always be split into complementary Pauli errors and commuted with the next round.
	Using this we can consider that after measuring \(X\) checks, only \(X\) errors happen, after \(Y\) checks only \(Y\) errors happen and after \(Z\) checks only \(Z\) errors happen.
	A decoder based on this assumption will not be optimal against more general noise but is sufficient for fault-tolerance.
	
	We follow the measurement schedule in Algorithm~\ref{alg:schedule}.
	At every step of the schedule a new value for the stabilizer of a given color is learned.
	These values are recorded and are the syndrome bits.
	Assuming the error model described above we observe the following relation between syndrome bits.
	An \(X\) error at time \(t\), just after the red (\(X\)) measurements, is going to flip the adjacent green (\(Y\)) stabilizer recorded at time \(t+1\) and the adjacent blue (\(Z\)) stabilizer recorded at time \(t+3\).
	Hence these two syndrome bits are linked in the syndrome graph.
	Doing the same for \(Y\) errors after green measurements and \(Z\) errors after blue measurements gives a syndrome graph with two disconnected components offset by one time step.
	Matching syndrome bits in this graph using minimum weight perfect matching gives the decoder \cite{hastings_dynamically_2021}.
	Note that partial magnetic, i.e. \(\bar{Z}\), logical operators, give non-trivial syndrome bits in only one of the two components of the syndrome graph. Whereas partial electric, i.e. \(\bar{X}\), logical operators, give nontrivial syndrome bits in both components of the syndrome graph.
	
	Measurement errors can also be tolerated since at each step any qubit is involved in at most one measurement.
	This makes it so that measurement errors are equivalent to local qubit errors.
	Indeed a measurement error is equivalent to an error on one of the qubits happening just before and just after the faulty measurement \cite{hastings_dynamically_2021}.
	
	\subsection*{Floquet codes can therefore be summarized as follows:}
	Given a planar graph \(\mathcal{G}=(V, E, F)\) tiling a surface with no boundaries, which is 3-valent and 3-colorable of the faces, we define a \(\llbracket n, k, d\rrbracket\) Floquet code with \(n=\vert V\vert\), \(k\) is defined as in Eq.~\eqref{eq:logdim} and \(d\) is a fraction of the minimal length of a homologically non-trivial cycle in the graph.
	This fraction depends on the size of the faces of the graph.
	The code space is stabilized by measuring gauge checks following a cyclic schedule of the three colors.
	In the steady state regime, at each step of the schedule the space is stabilized by the face stabilizers and all the gauge checks of the last measured color.
	This space is Clifford equivalent to the space stabilized by a 2D homological code as detailed in Table~\ref{tab:homologicaleq} and drawn in Figure~\ref{fig:genericsteps}.
	Decoding can be performed by matching syndrome bits with minimum weight perfect matching.
	
	\subsection{Boundaries and Corners for Floquet Codes}
	\label{sec:bound}
	
	One natural way to introduce boundaries to a 2D surface is to remove from \(F\) some of the face.
	This creates boundaries which are called colored boundaries for 2D color codes.
	The vertices belonging to these boundaries are still 3-valent.
	The edges belonging to these boundaries form bi-colored cycles which is an issue raised in \cite{hastings_dynamically_2021}.
	Indeed, as we just saw, these cycles are measured when successively measuring these two colors.
	This is unsurprising since the set of faces \(F\) is actually irrelevant for the definition of the Floquet code.
	Indeed it was convenient to derive its properties but removing some faces or even completely changing \(F\) does not change what measurements are done at what time.
	In turn it does not change the stabilizers that stabilizes the space at each step of the measurement schedule.
	So removing a face from \(F\) cannot change the fact that the bordering cycle of edges will be measured.
	In \cite{hastings_dynamically_2021} the authors propose that non-uniform and more clever measurement schedules could be used to escape this problem but also show some obstructions.
	
	Color codes have a rich theory of boundaries, corners and twists \cite{kesselring_boundaries_2018}.
	Taking inspiration from them we introduce corners, which are points where different colored boundaries meet.
	Said differently we also allow vertex punctures in the graph which removes some vertices and edges.
	Structurally a corner is a vertex which belongs to only one face and only two edges.
	The two missing faces are two boundaries of different colors.
	At each corner we associate the same color as the face it belongs to and a single-qubit gauge check of this color.
	We also refer to them as corner checks.
	
	Note that inner logical operators cannot terminate at colored boundaries.
	When terminating them at some corners they commute with every gauge checks except with the corner checks where they terminate.
	This means that such operators represent logical operators at the steps where the corner checks at their ends are not measured.
	
	Boundaries between corners of the same color have an even number of qubits whereas boundaries between corners of different colors have an odd number of qubits.
	This difference is crucial as a Pauli operator on an odd number of qubits cannot be measured in two steps by measuring separately the two other complementary Pauli operators on these qubits.
	On the contrary this is possible with an even number of qubits.
	This means that a boundary between two corners of the same color should not be considered as a boundary as it could be measured eventually\footnote{If the measurement schedule is such that the color of the corners is measured right before the color of the other edges in the boundary, then the boundary will be measured directly. If the schedule measures the other color just before the color of the corners, then a two-qubit error (entangling the corners) is sufficient for accidentally measuring the boundary. Even if the corners are far apart this is unwanted behavior for an error correcting code.}.
	In fact it should be considered as a stabilizer and its two corners should be linked into a single gauge check.
	On the contrary, boundaries between corners of different colors will not be measured by the schedule of gauge check measurements, that is because they have odd length.
	These are the proper boundaries to introduce in Floquet codes.
	These two possibilities are shown in Figure~\ref{fig:evenoddbound}.
	\begin{figure}[ht!]
		\centering
		\includegraphics[height=.05\textheight]{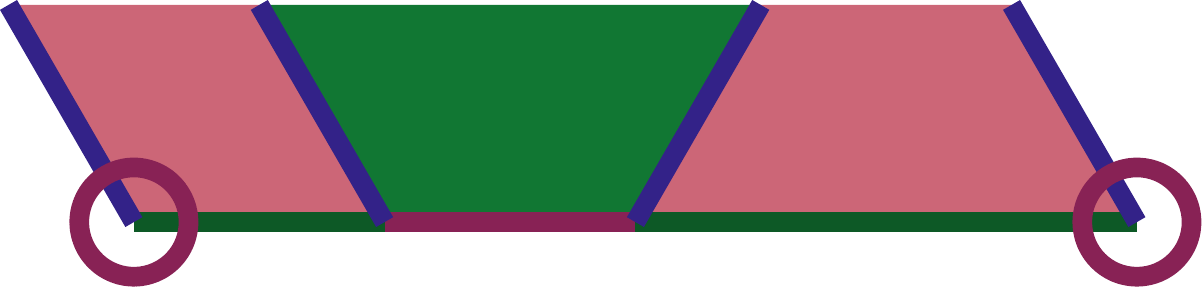}\hfil
		\includegraphics[height=.05\textheight]{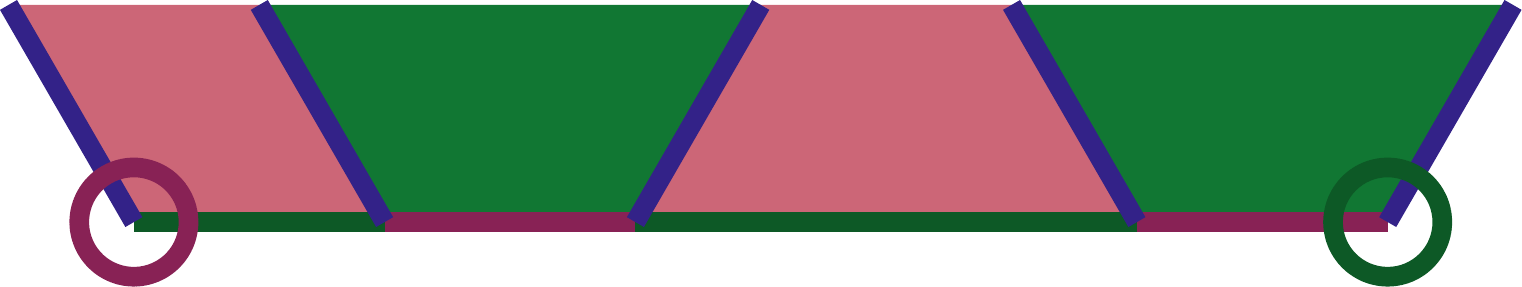}
		\caption{Examples of even and odd blue boundaries.
			On the left a blue boundary joining two red corners.
			It has an even length and should therefore be considered as a stabilizer.
			On the right a blue boundary joining a red corner and a green one.
			It has odd length and will not be measured by the schedule of measurements.}
		\label{fig:evenoddbound}
	\end{figure}
	
	Looking at the Floquet code in the equivalent homological code picture we can assign smooth and rough labels to the boundaries.
	Right after having measured the color \(c\), the boundaries of color \(c\) are smooth and the other two colors are rough, see for instance in Figure~\ref{fig:stdschedule}.
	With this we can see that the layout of 2D triangular color codes will not make Floquet codes with a non-zero number of logical qubits.
	But making an hexagonal patch with 6 boundaries alternating the three colors cyclically twice allows to make exactly one logical qubit.
	Indeed it will have two rough boundaries alternating with two smooth boundaries \cite{bravyi_quantum_1998}.
	In Figure~\ref{fig:planarfloquet} we have drawn two examples, one based on the hexagonal lattice and the other on the square-octagon lattice.
	For aesthetics we mostly use the hexagonal layout but the square-octagon one is also valid and may even be less resource intensive as the square-octagon lattice is more qubit efficient.
	Moreover any color code layout with the same corner and boundary configuration would also work.
	The overall performance will be a combination of qubit efficiency of the lattice and decoder performance on it.

	\section{Planar Floquet Codes}
	\label{sec:planarfloquetcodes}
	
	\subsection{Definition}
	To define planar Floquet codes we use the layout of a 2D color codes on a disk with colored boundaries such that all boundaries have odd length.
	This requirement translate into the fact that the corners must cycle through the three colors and in turn that the boundaries must also cycle through the 3 colors.
	We can do the same computation as in Section~\ref{sec:floquetfromcolorcodes}, denoting the number of corners as \(n_c\), we have
	\begin{equation}
	n_g = n_e + n_c -1.
	\end{equation}
	A disk has trivial homology and inner logical operators cannot terminate at boundaries or corners, besides there are no linear dependencies between the faces because of the boundaries so
	\begin{equation}
	n_s = n_f.
	\end{equation}
	Vertices are still three valent when counting the corner checks so
	\begin{equation}
	3n_v = 2n_e + n_c.
	\end{equation}
	Finally a disk has Euler characteristic 1
	\begin{equation}
	n_f - n_e + n_v = 1.
	\end{equation}
	Counting the number of logical qubits as a subsystem code also yields zero:
	\begin{equation}
	n_L = n_v - \frac{n_g-n_s}{2} - n_s = \frac{2n_v - n_e - n_c + 1 - n_f}{2} = 0.
	\end{equation}
	
	A planar Floquet code with \(3(k+1)\) boundaries has \(k\) logical qubits.
	This can be checked by looking at the equivalent homological code at any given time step.
	The boundaries of two of the colors will correspond to rough boundaries and the third color to smooth boundaries.
	There will therefore be \((k+1)\) smooth boundaries alternating with \((k+1)\) rough boundaries which is known to host \(k\) logical qubits \cite{delfosse_generalized_2016}.
	
	The simplest example is then an hexagonal patch with six boundaries cycling through colors \((R,G,B,R,G,B\)).
	The patch is hexagonal but the bulk can have any color code structure, such as one like in Figure~\ref{fig:generic}.
	As a color code such system has 4 logical qubits but as a Floquet code it has a single logical qubit.
	\begin{figure}[ht!]
		\centering
		\includegraphics[width=.08\linewidth]{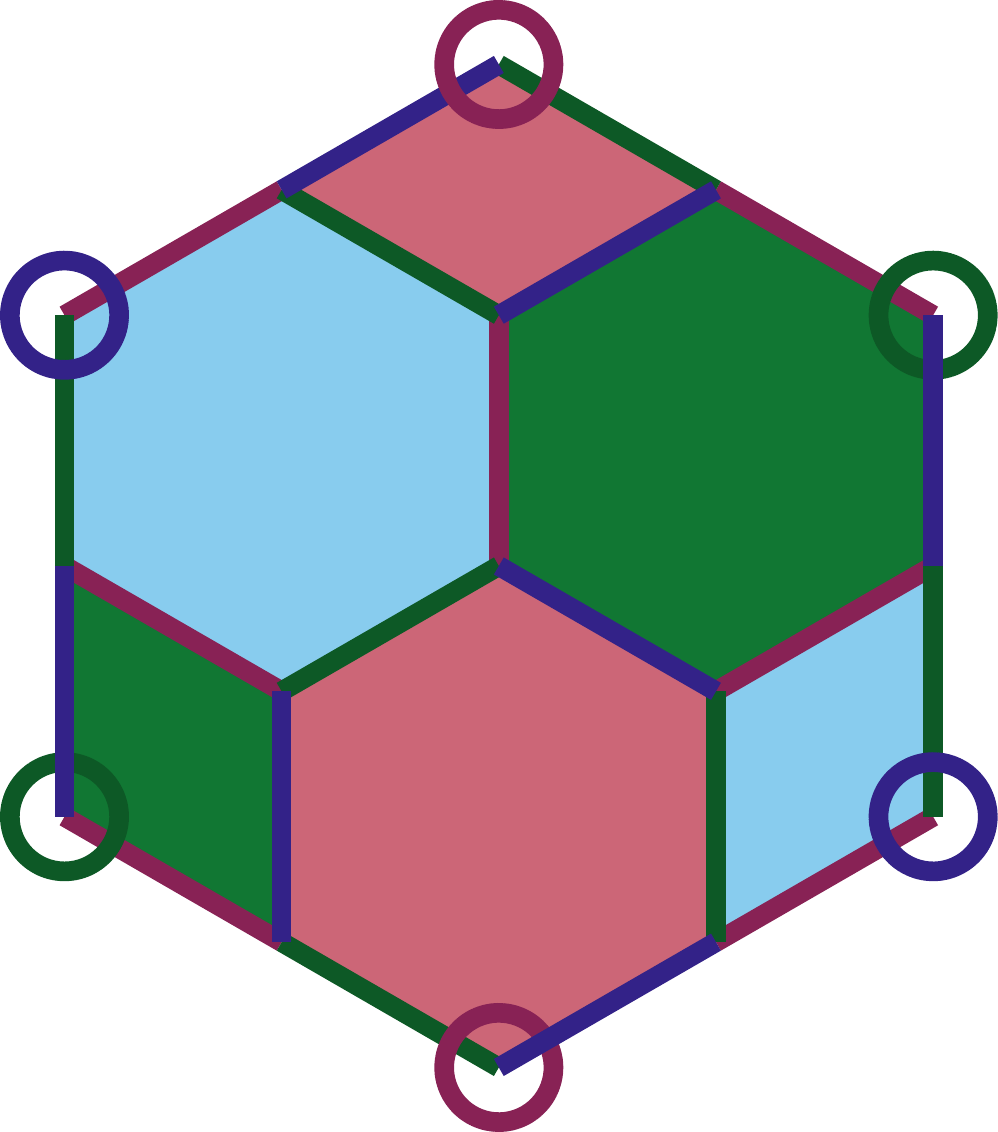}\hfil
		\includegraphics[width=.2\linewidth]{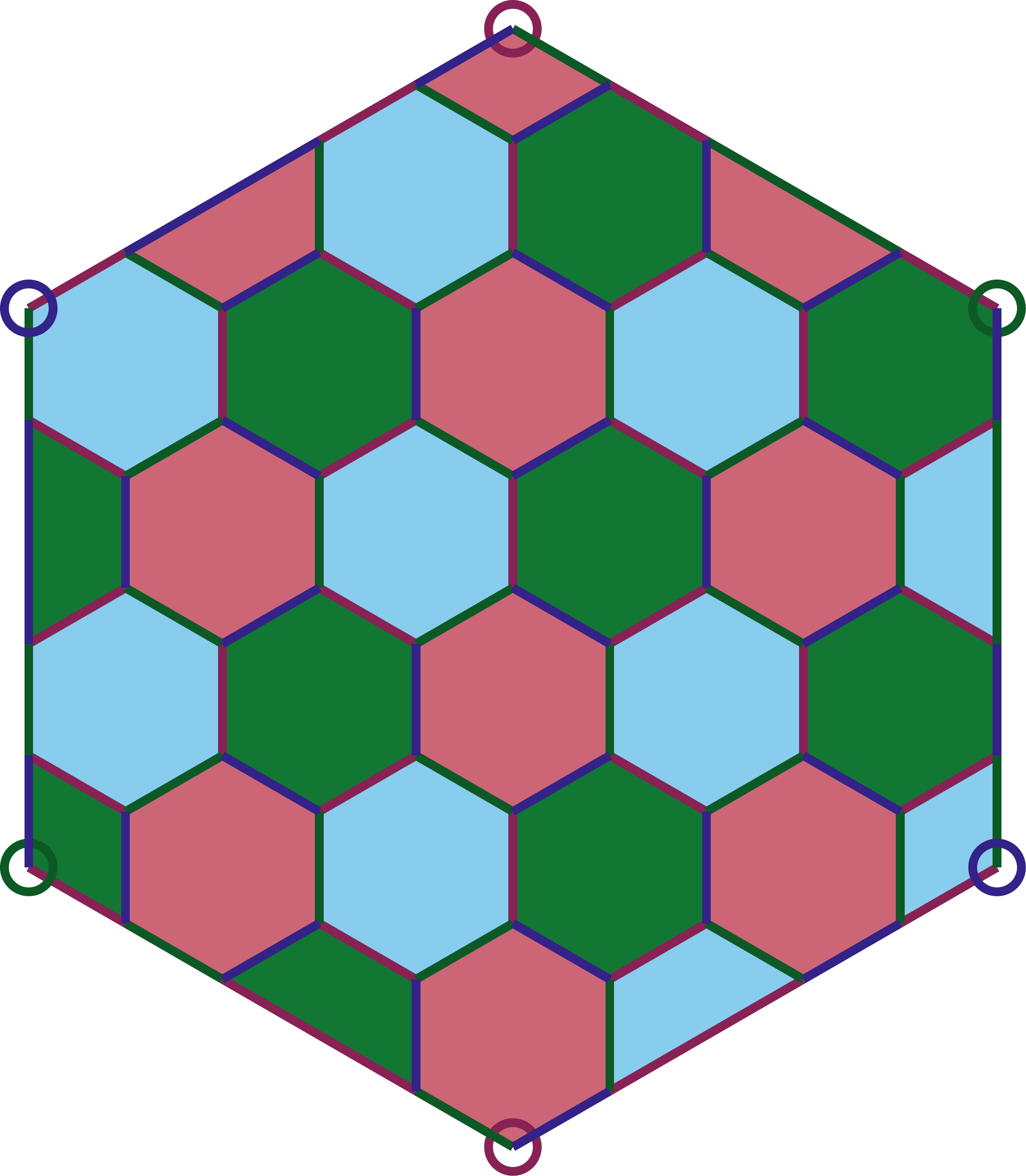}\hfil
		\includegraphics[width=.15\linewidth]{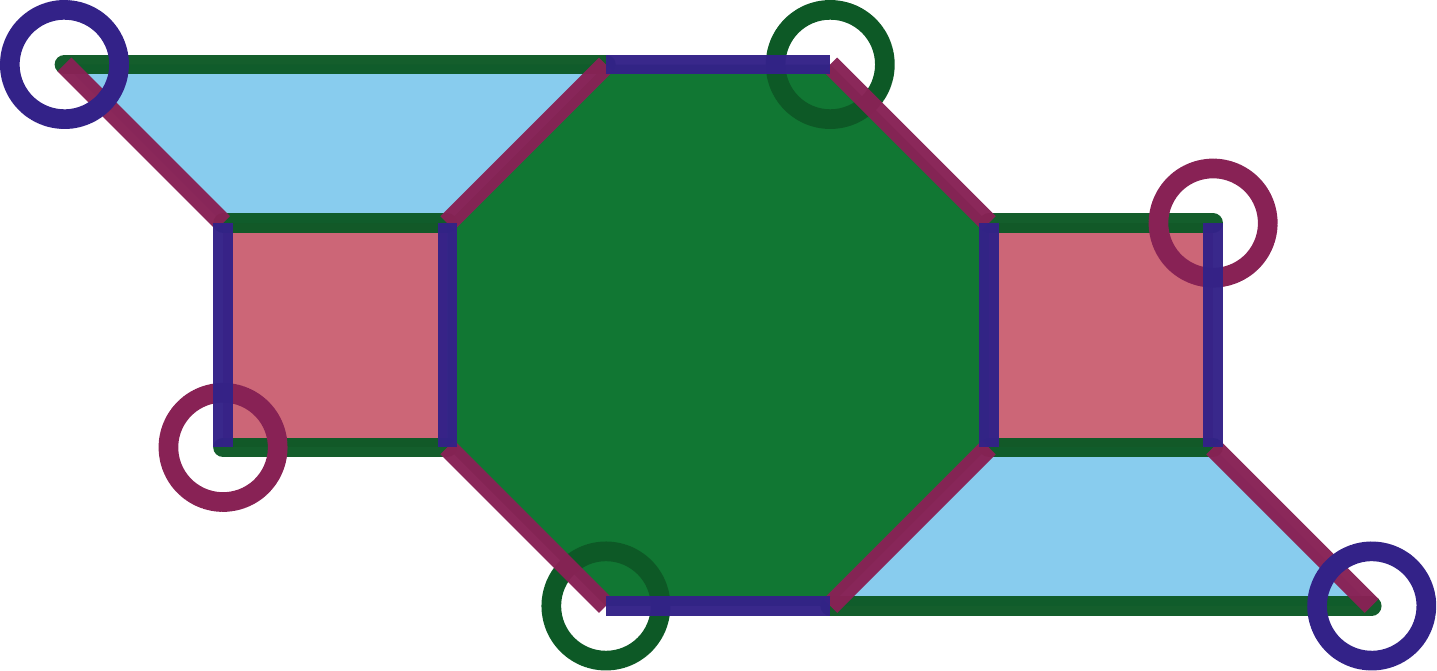}\hfil
		\includegraphics[width=.32\linewidth]{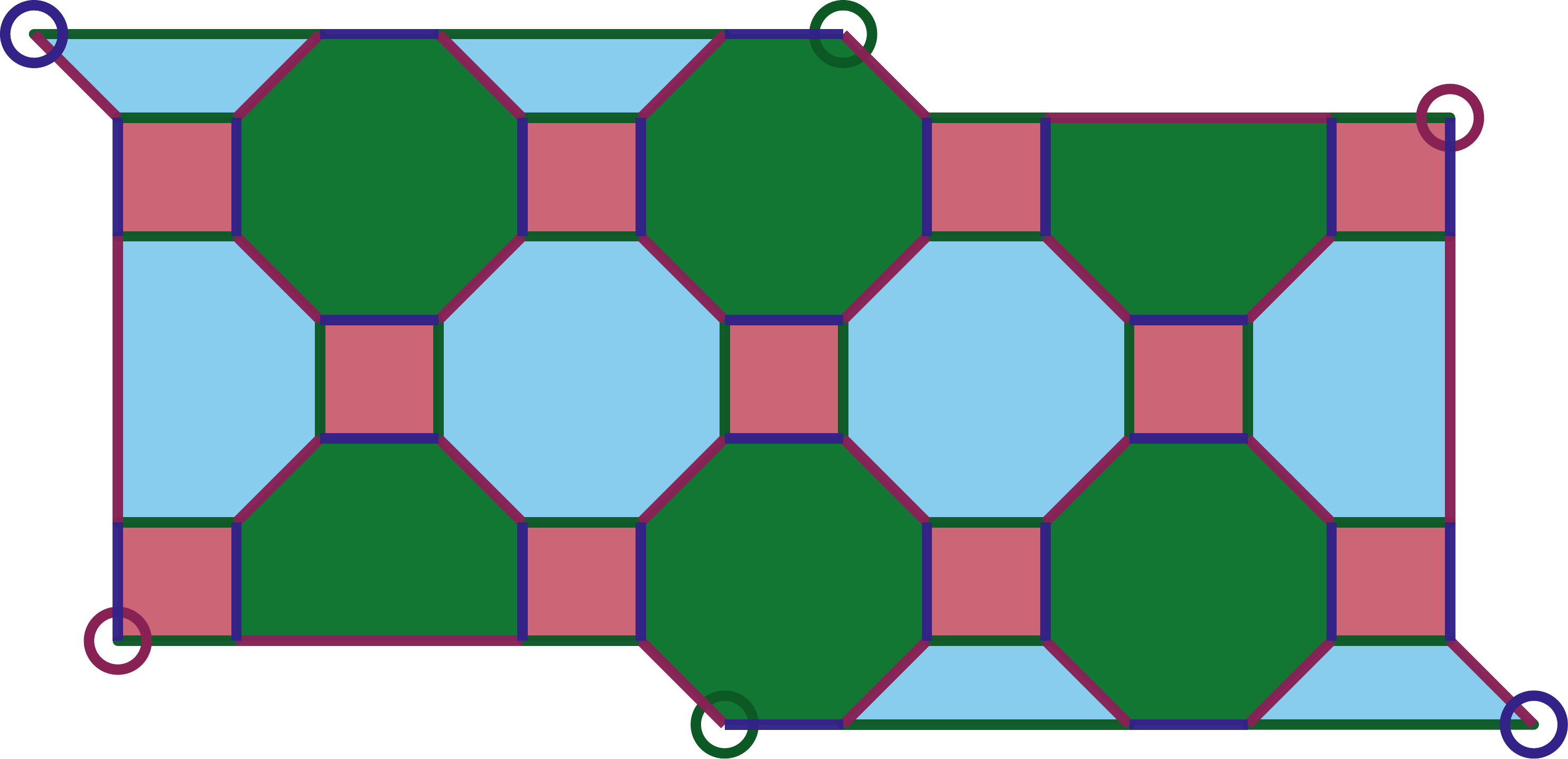}
		\caption{Layouts for planar Floquet logical qubits based on the hexagonal and square-octagon lattices.
		Each vertex hosts a physical qubit.
		Each edge or circle is a gauge check.
		The faces are stabilizers and there are no inner logical operators.
		Products of checks ending at corners represent logical operators at rounds where the corners they end on are not measured.
		For both lattices we show first two smallest instances.}
	\label{fig:planarfloquet}
	\end{figure}

	\subsection{Dynamics}
	We can examine the dynamics of the logical space of a single qubit planar Floquet code.
	The cyclic sequence of measurements is exemplified in Figure~\ref{fig:stdschedule}.
	We see that the hexagonal patch is gradually rotating counter-clockwise when considering the geometry of the equivalent homological code.
	\begin{figure}[ht!]
	\centering
	\includegraphics[width=.3\linewidth]{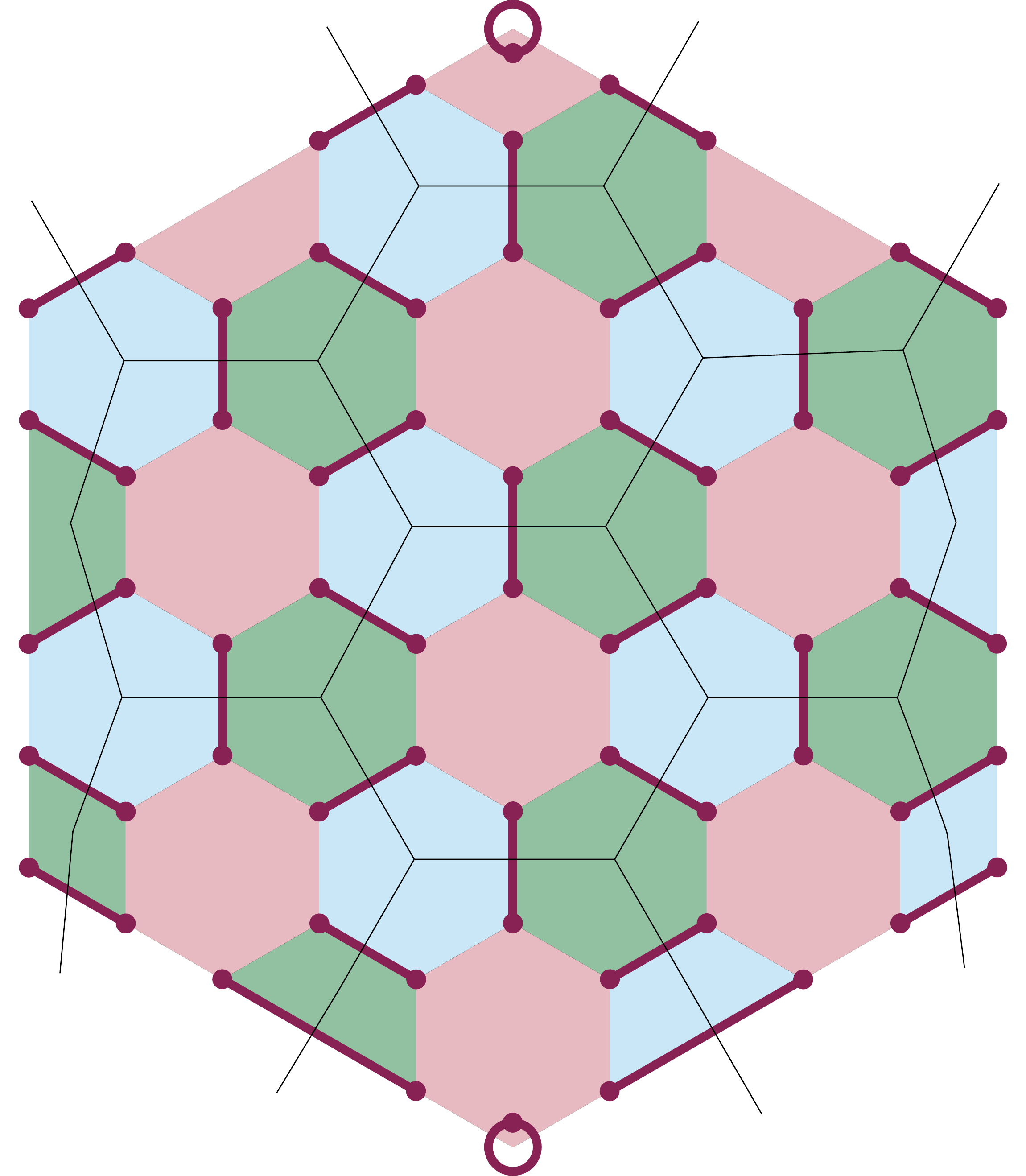}\\\(\swarrow\qquad\qquad\qquad\qquad\nwarrow\)\\\vspace{-1em}
	\includegraphics[width=.3\linewidth]{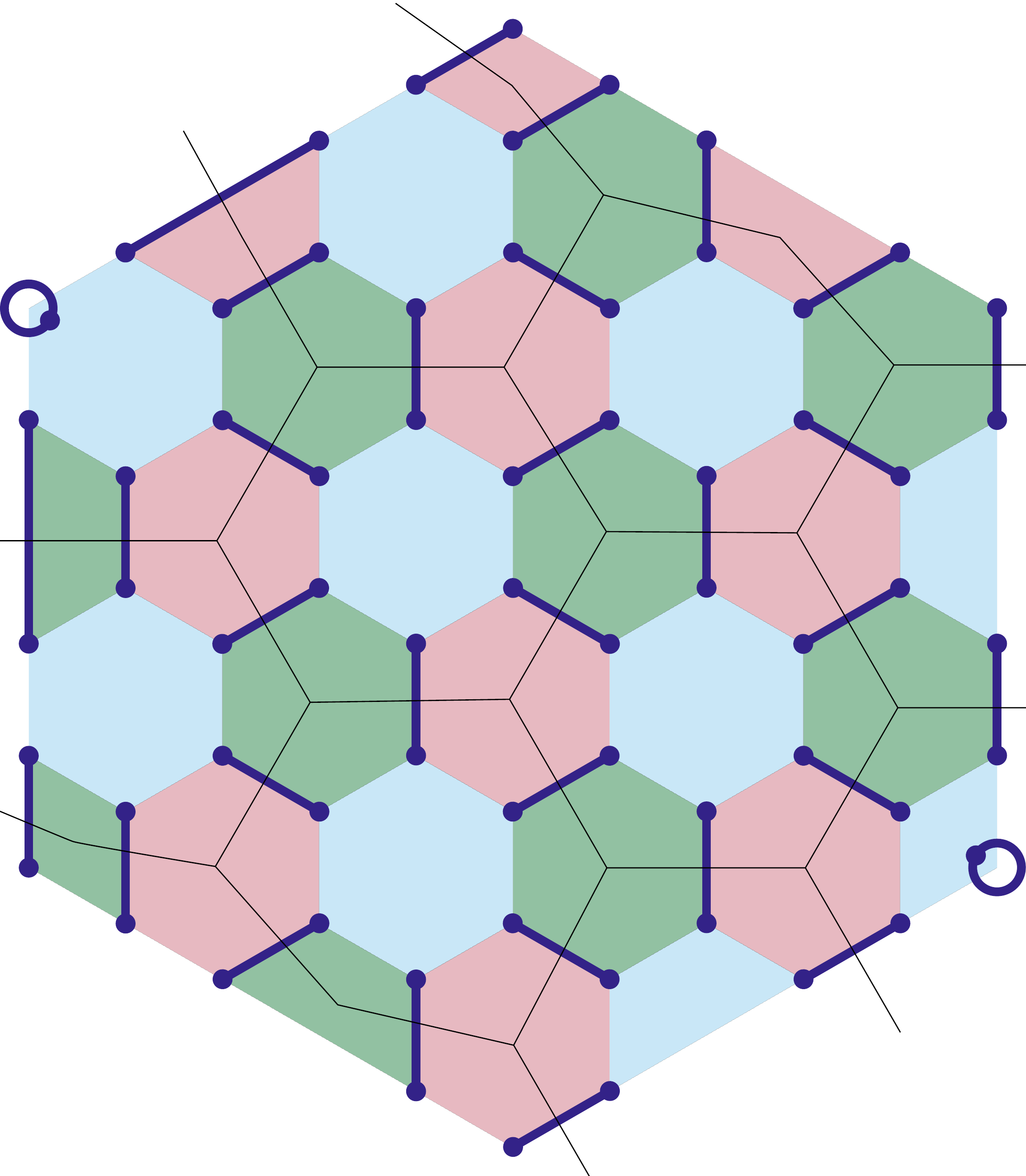}\hfil\raisebox{2.5cm}{\(\rightarrow\)}\hfil
	\includegraphics[width=.3\linewidth]{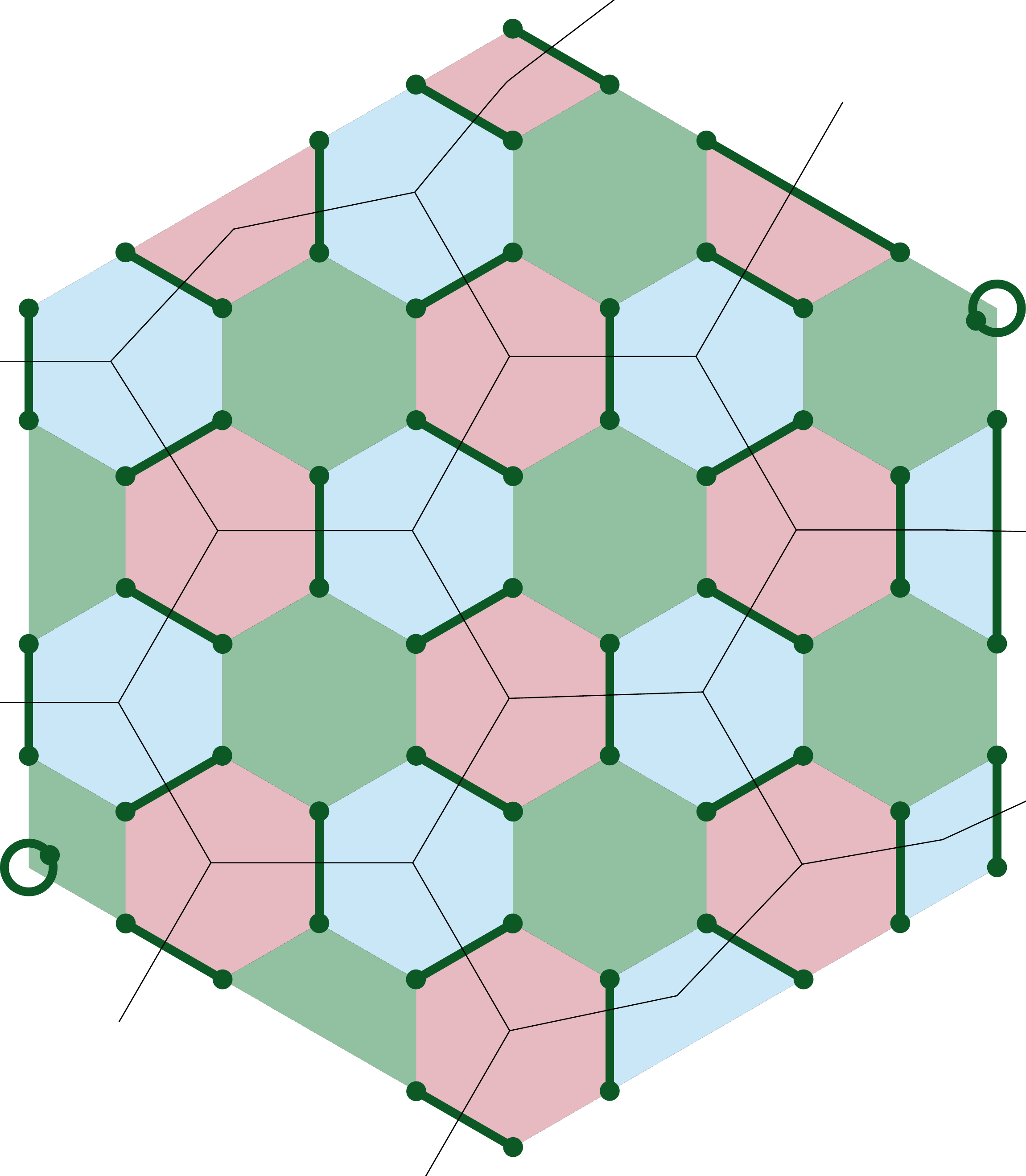}
	\caption{Representation of the schedule of measurements ---red (\(X\)), blue (\(Z\)), green (\(Y\)) and repeat--- and how the system evolves.
	At each time step the Clifford equivalent surface code is drawn in thin black edges.
	It has two rough boundaries facing two smooth boundaries which are rotated counter-clockwise by 60 degrees at each step.}
	\label{fig:stdschedule}
	\end{figure}

	In fact, if we track the evolution of logical operators they gradually rotate clockwise and are mapped from smooth to rough at every time step.
	We show in Figure~\ref{fig:logicaldynamics} the evolution starting from a smooth logical operator at step 0.
	The way to track the evolution is to apply the standard Gottesman-Knill rule \cite{gottesman_heisenberg_1998} to track Pauli operators evolution under Pauli measurements.
	At each time step we can multiply the current logical operator by some currently measured gauge checks to make it commute with the next set of gauge checks.
	This alternatively make the logical operator grow onto the next boundary then shrinks it in a caterpillar-like motion going clockwise.
	Moreover the logical operator is alternating between rough and smooth boundaries and as the measurement cycle has odd length a Hadamard is applied every three steps to the logical qubit.
	
	\begin{figure}[ht!]
	\centering
	\includegraphics[width=.3\linewidth]{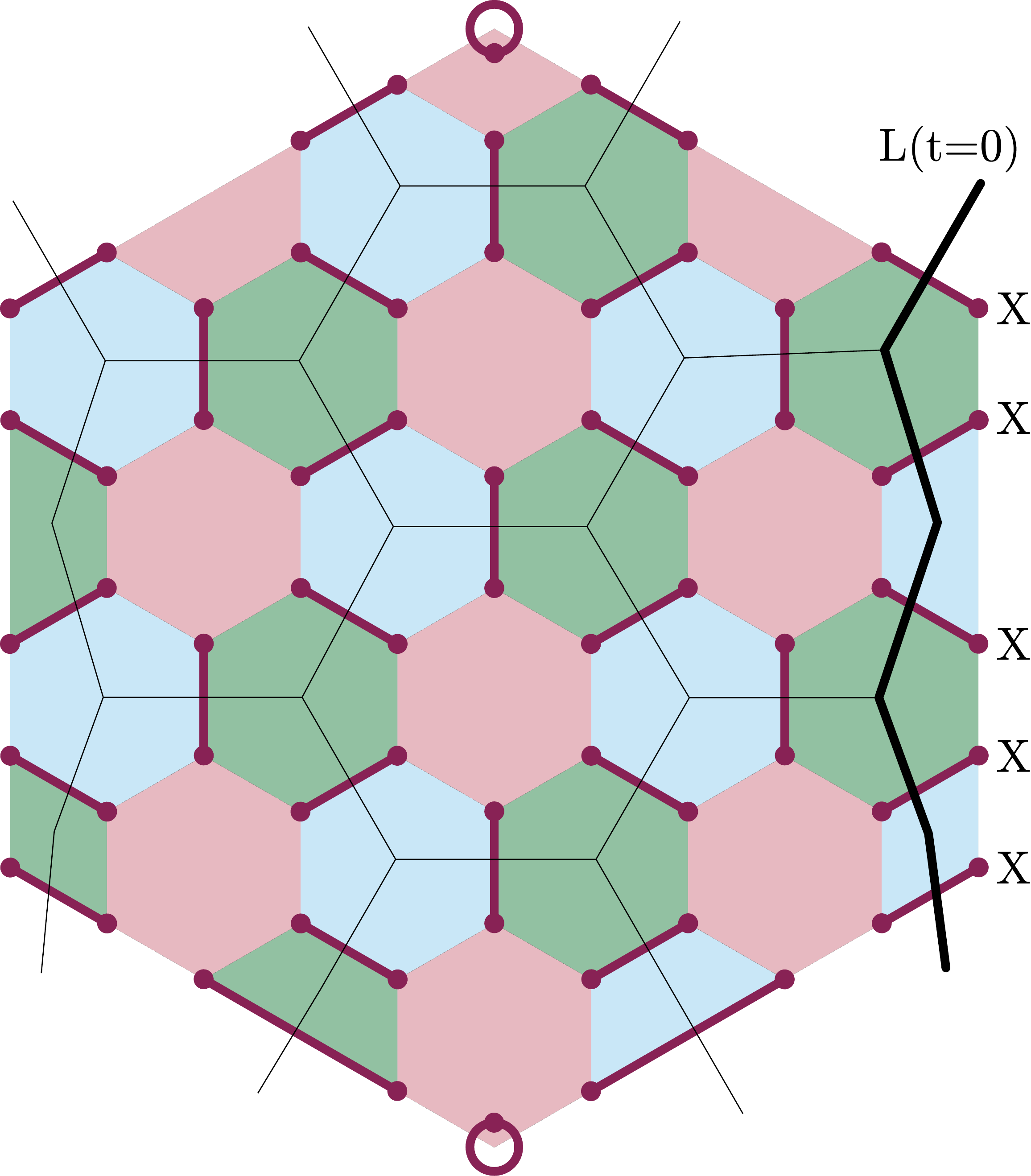}\hfil\raisebox{2.5cm}{\(\rightarrow\)}\hfil
	\includegraphics[width=.34\linewidth]{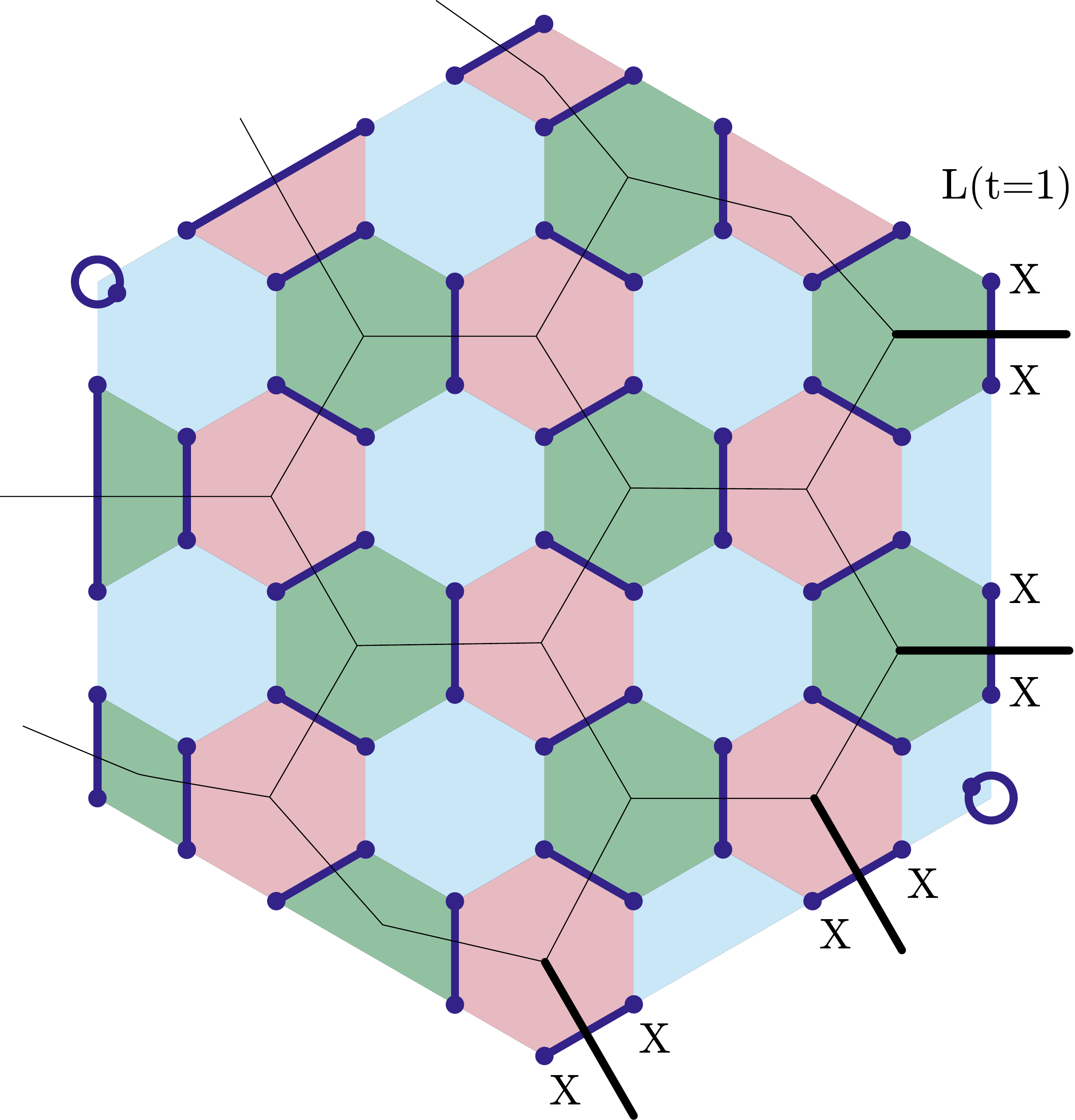}\\[1em]\(\qquad\qquad\qquad\qquad\qquad\qquad\qquad\qquad\qquad\qquad\downarrow\)\\[1em]
	\includegraphics[width=.3\linewidth]{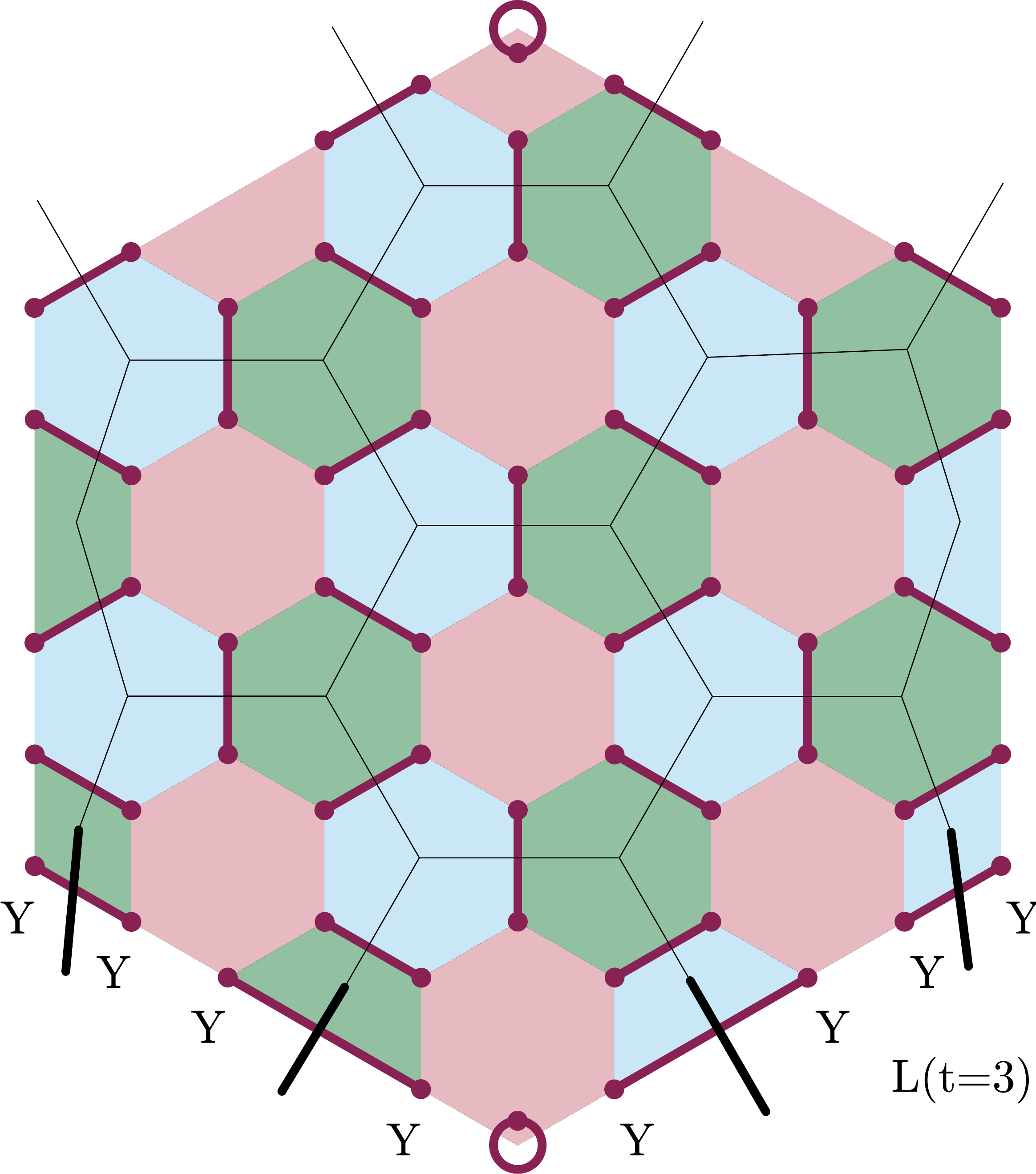}\hfil\raisebox{2.5cm}{\(\leftarrow\)}\hfil
	\includegraphics[width=.36\linewidth]{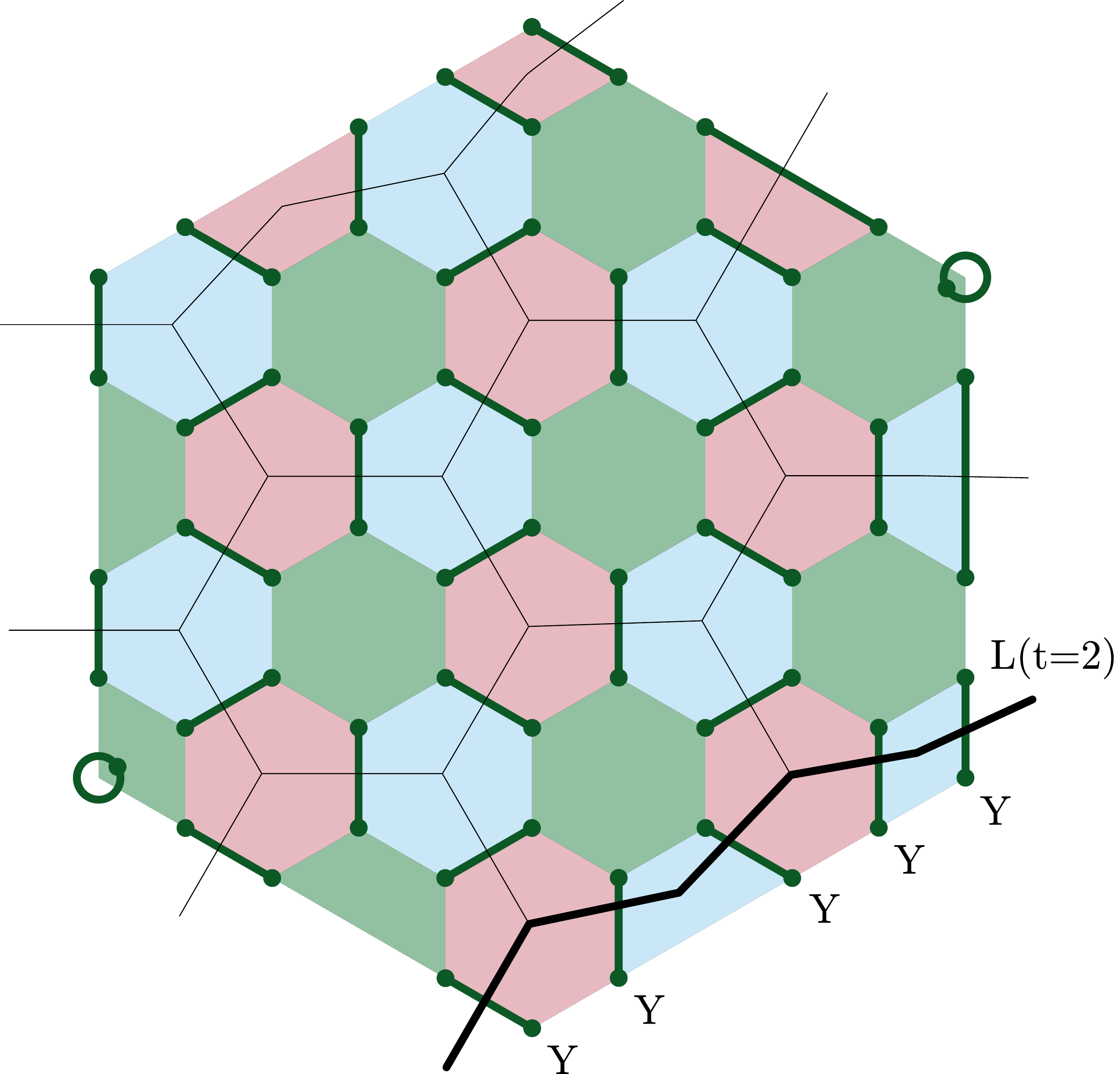}
	\caption{The dynamical evolution of logical operators in a planar Floquet code.
			 We highlight a starting logical operator, right after measured red (\(X\)) checks, and track its evolution.
			 Going from a logical operator of the equivalent homological code to the one of the Floquet code can be done using Table~\ref{tab:homologicaleq}.
			 In order to go from one step to the other we multiply the current logical operator by some of the current gauge checks so that it commutes with the next gauge checks.
			 This can be used to change its Pauli type and make it grow along an adjacent boundary.
		 	 For some steps part of the logical operator is measured by the next gauge checks (but never the whole operator) which we use to shrink it.
	 	 	 Overall the logical operator is alternating between smooth and rough at every step and is gradually moving clockwise along the boundary.
 	 	 	 The net effect of a cycle of three steps is to map smooth to rough and rough to smooth.
 	 	 	 Using the convention that smooth boundaries correspond to \(Z\)-logical operators and rough ones to \(X\)-logical operators implies that a Hadamard is applied to the logical qubit.}
  	 	 \label{fig:logicaldynamics}
	\end{figure}
	
	\subsection{Constant Size Logical Operators}
	\label{sec:constdistance}
	A previous version of this manuscript was erroneously implying that the planar Floquet codes obtained this way could be used for fault-tolerant quantum computation.
	It is in fact not the case as there exist constant size logical operators.
	
	Schematic examples for such space-time operators are represented in Fig.~\ref{fig:constanterror} and \ref{fig:constanterroranticomm}.
	
	Considering that the logical operators are rotating clockwise, one can have an electric particle emerging at the tail (one goes clockwise from tail to head) of a rough boundary. 
	Two steps later the opposite rough boundary has rotated towards it and the particle can be absorbed in it.
	On can then check that any space-time sheet representing a logical operator of the opposite type is crossed an odd number of time by this error path and creates therefore a logical error.
	The error can stay at a constant distance from the corner and has to stay only a constant number of steps in the bulk so there are constant-size errors independent of the size of the code patch.
	
	\begin{figure}[ht]
		\centering
		\includegraphics[width=.6\linewidth]{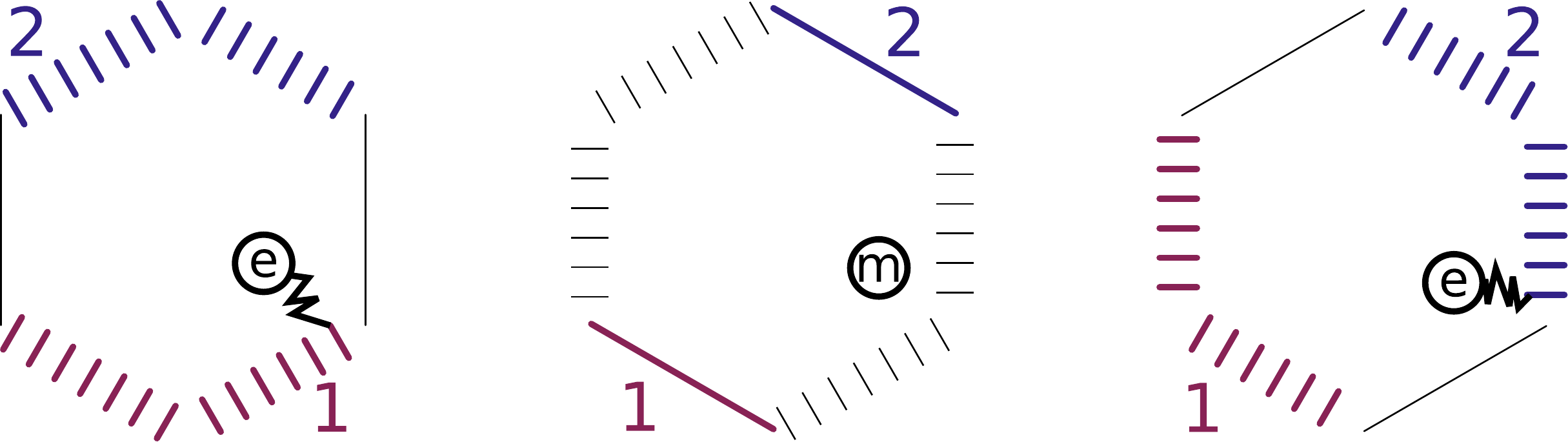}
		\caption{A small space time error in three steps which is undetectable and non-trivial. The code is just schematically represented by rough and smooth boundaries of the effective surface code present at each time step.
			The error consists in creating an electric excitation from a rough boundary right next to an adjacent smooth boundary. Then wait the next time step}
		\label{fig:constanterror}
	\end{figure}
	
	\begin{figure}[ht]
		\centering
		\includegraphics[width=.6\linewidth]{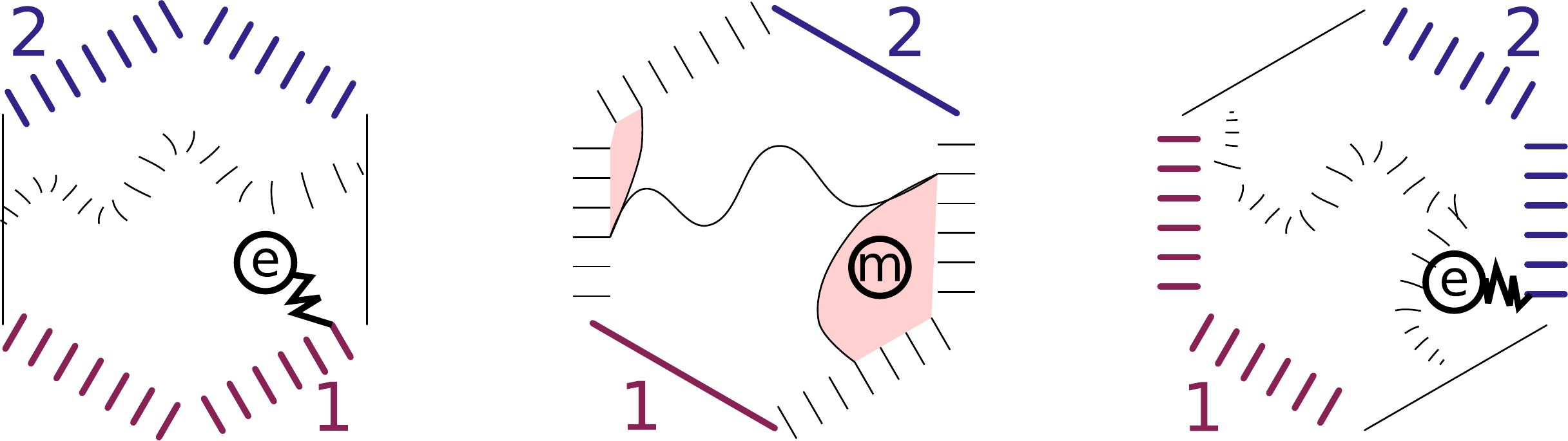}
		\caption{Tracking a generic logical operator which is flipped by the error. The red regions show where the operator has to be multiplied by checks and stabilizers in order to stay attached to the correct type of boundaries. 
		One can see that space time sheet supporting the logical operator has to pass by the left right corner and therefore cross once the error path.}
		\label{fig:constanterroranticomm}
	\end{figure}

	One can note that the same type of error path emerging from the head of the rough boundary and going clockwise into the next boundary would actually be equivalent to a stabilizer and harmless to the logical information.
	
	\subsection{Decoding}
	Decoding can be done in the same way as it is done when there are no boundaries.
	The only difference is that one can match some non-trivial syndrome bits with some of the boundaries in the syndrome graph.
	Recall from Section~\ref{sec:decoder} that errors creating non-trivial syndrome bits within the same connected component of the syndrome graph correspond to partial magnetic, i.e. \(\bar{Z}\), logical operators.
	This means that they can terminate at rough boundaries.
	The rough boundaries are rotating with the dynamics of the system so they form spiraling surfaces in the shape of a DNA strand in the space-time syndrome graph.
	
	Another way of seeing the constant distance is as follows: Since the rough boundaries are spiraling and making a full revolution in a constant number of steps, in the space-time decoding graph there are constant size space-time paths from one rough boundary to the opposite one, see Sec.~\ref{sec:constdistance}.
	
	\subsection{Computation}
	In this section we investigate how to perform quantum computation on the logical qubits protected by planar Floquet codes.
	We show how to perform Hadamard and CNOT gates using the natural dynamics of the system, and code deformation techniques.
	Unfortunately there are no natural \(S\) gate, but a few number of magic states for the \(S\) gate are sufficient to complete the set of Clifford gates.
	In order to add \(T\) gates for universal quantum computation standard magic state distillation methods can be used.
	
	\subsubsection{Hadamard}
	We have seen from the dynamics of the logical operators that the Hadamard gate is naturally applied to the logical qubits.
	More precisely every three steps the rough logical Pauli is mapped to the smooth one.
	In order to apply Hadamards differentially to some of the logical qubits and not the others, we can simply de-synchronize the qubits needing a Hadamard from the other ones.

	For instance if one repeat each schedule step twice for a given logical qubit, in 6 steps it will be back to its initial subspace but with a logical Hadamard applied on it.
	In the same time other logical qubits changing measurement at each step will have undergone two Hadamard gates, i.e. they are back to their initial state.
	
	Repeating the measurements to (de-)synchronize patches will unfortunately allow for some type of errors to accumulate a bit which can decrease the error correction performance a bit.
	Since the slow down in measurements is by a constant factor this is still fault-tolerant.
	
	We can also play with the synchronization to rotate qubits with respect to one another in order to targeted some specific configurations and choose which boundary is facing which.
	
	\subsubsection{CNOT gates}
	To perform a CNOT gate we can once again use the dynamics of the system to our advantage.
	Consider a nonagon planar Floquet code, i.e. with nine boundaries: it encodes two logical qubits.
	We schematically represent such a patch characterized by its pattern of boundaries and corners in Figure~\ref{fig:9sides}.

\begin{figure}[ht]
	\centering
	\includegraphics[width=.2\linewidth]{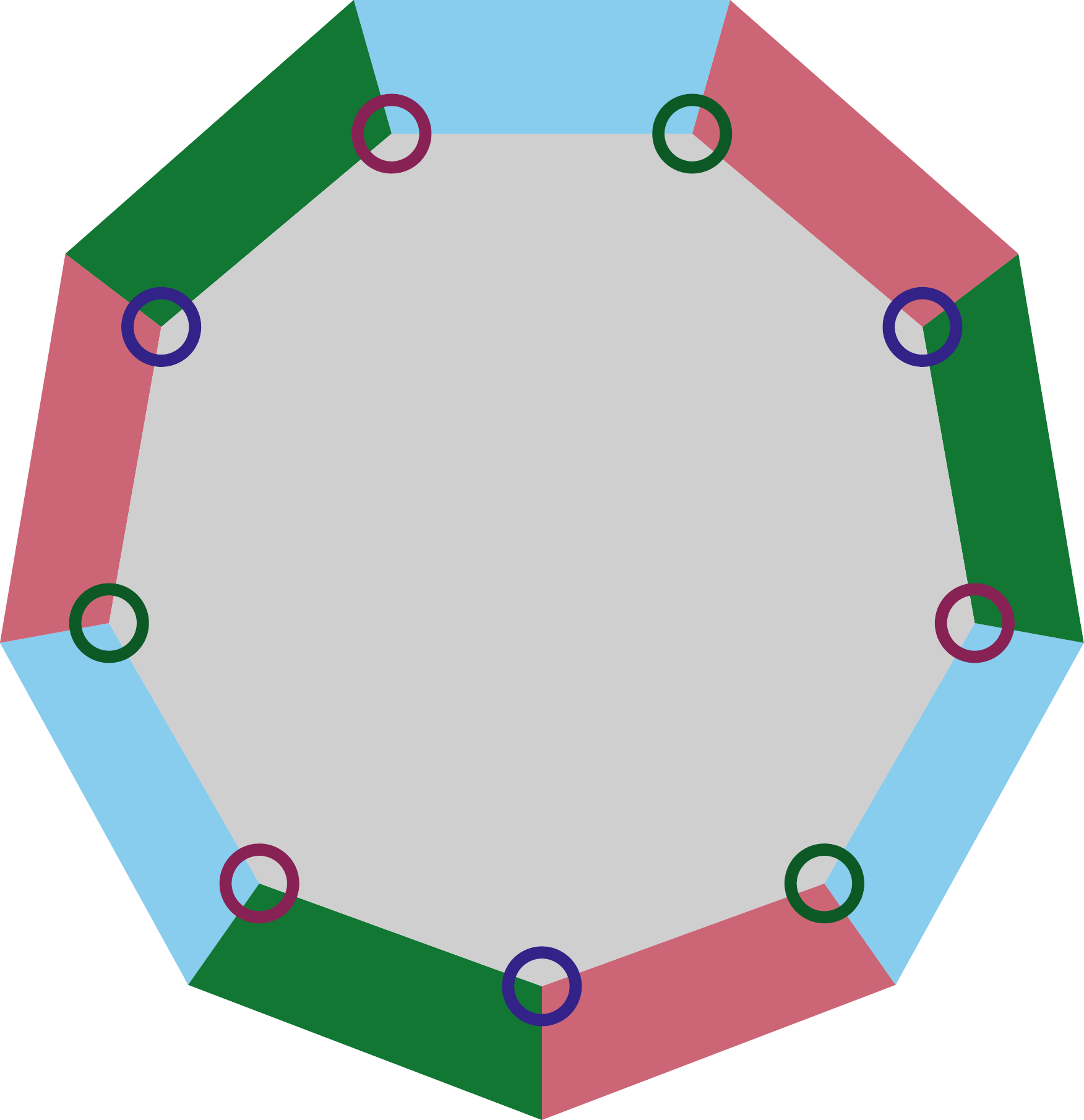}\hfill
	\includegraphics[width=.2\linewidth]{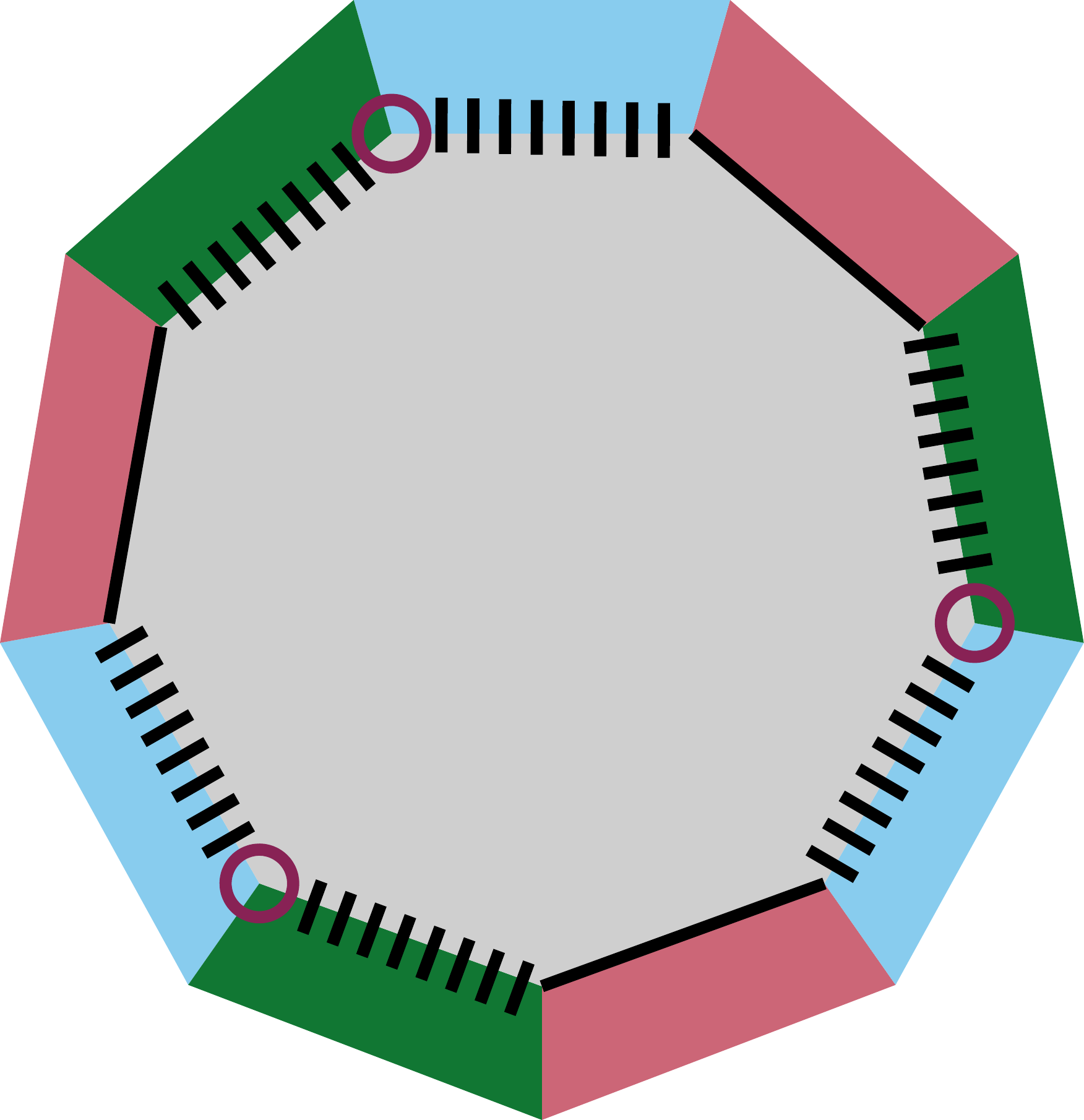}\hfill
	\includegraphics[width=.2\linewidth]{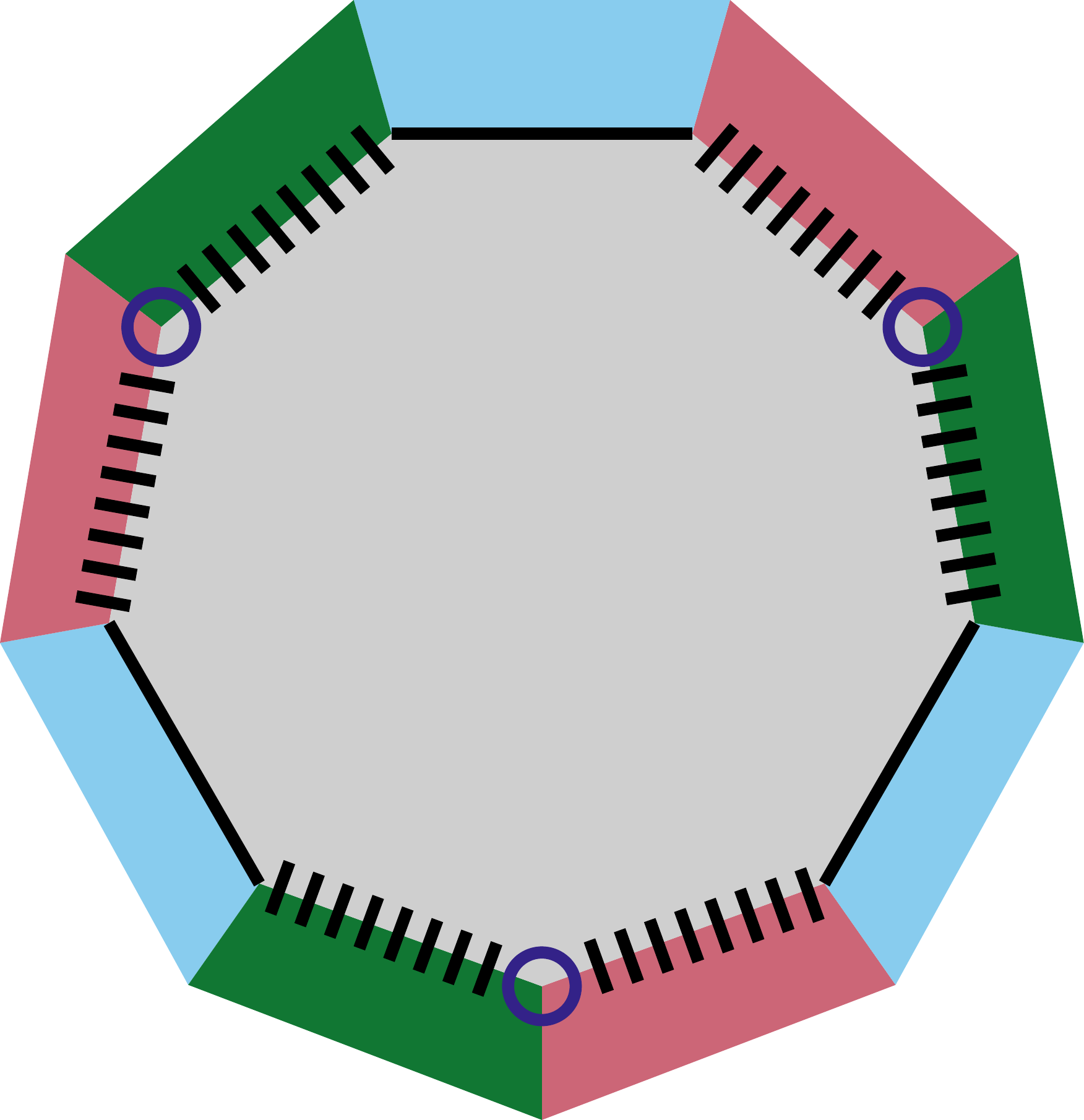}\hfill
	\includegraphics[width=.2\linewidth]{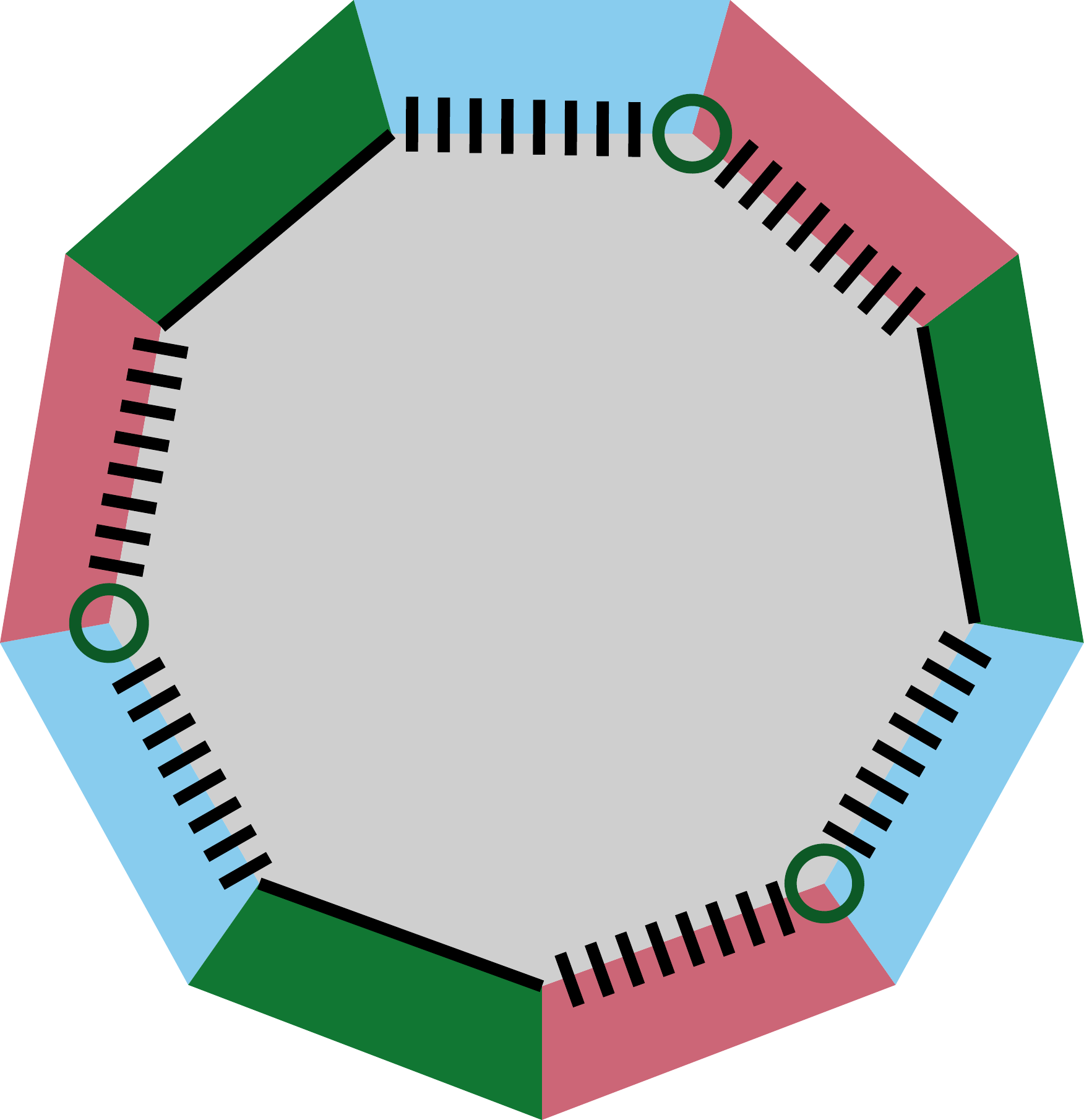}\hfill
	\caption{Schematic representation of a two-qubits planar Floquet patch with the three different configurations in time with the equivalent rough and smooth homological code boundaries.}
	\label{fig:9sides}
\end{figure}

	We have seen in the previous section that logical operators gradually move clockwise around the patch in a sort of ``laridé'' dance\footnote{A traditional dance from Brittany where participant form a circle and go around holding each other's hands.}.
	This is still the case when the patch hosts more than one logical qubit.
	The crucial difference, in the case there are more than one logical qubit, is that some boundaries then represent joint logical operators so the dynamics of the system is to naturally entangle the logical qubits at certain time steps.
	We represent this dance in Figure~\ref{fig:laride}.
	After three time steps the logical information has undergone two Hadamard gates and a CNOT.
	After nine steps it would have undergone two Hadamard and a SWAP.
	After 18 steps it would be back to its initial state.
		\begin{figure}[ht!]
		\centering
		\includegraphics[width=.2\linewidth]{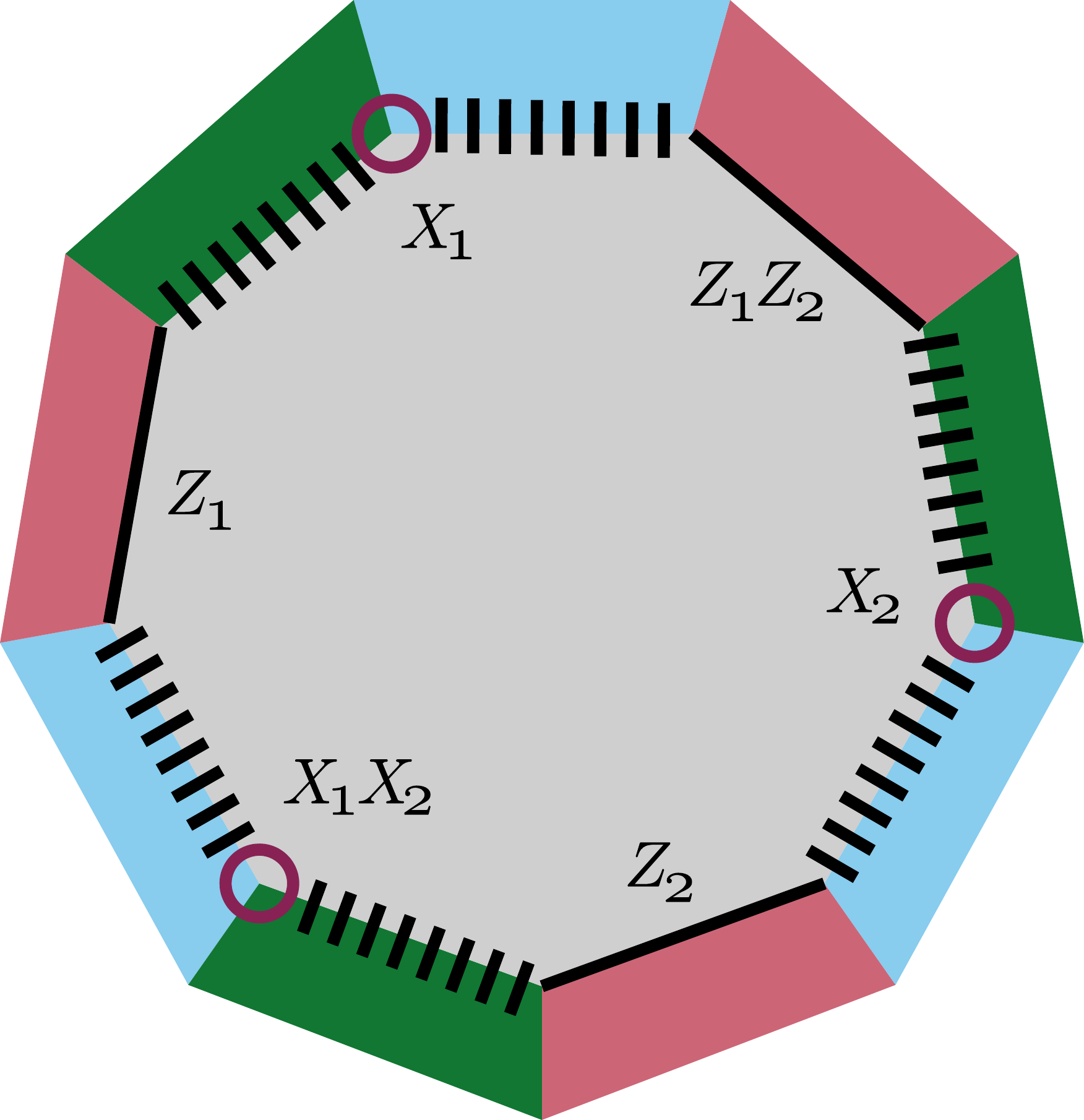}\hfil\raisebox{1.5cm}{\(\rightarrow\)}\hfil
		\includegraphics[width=.2\linewidth]{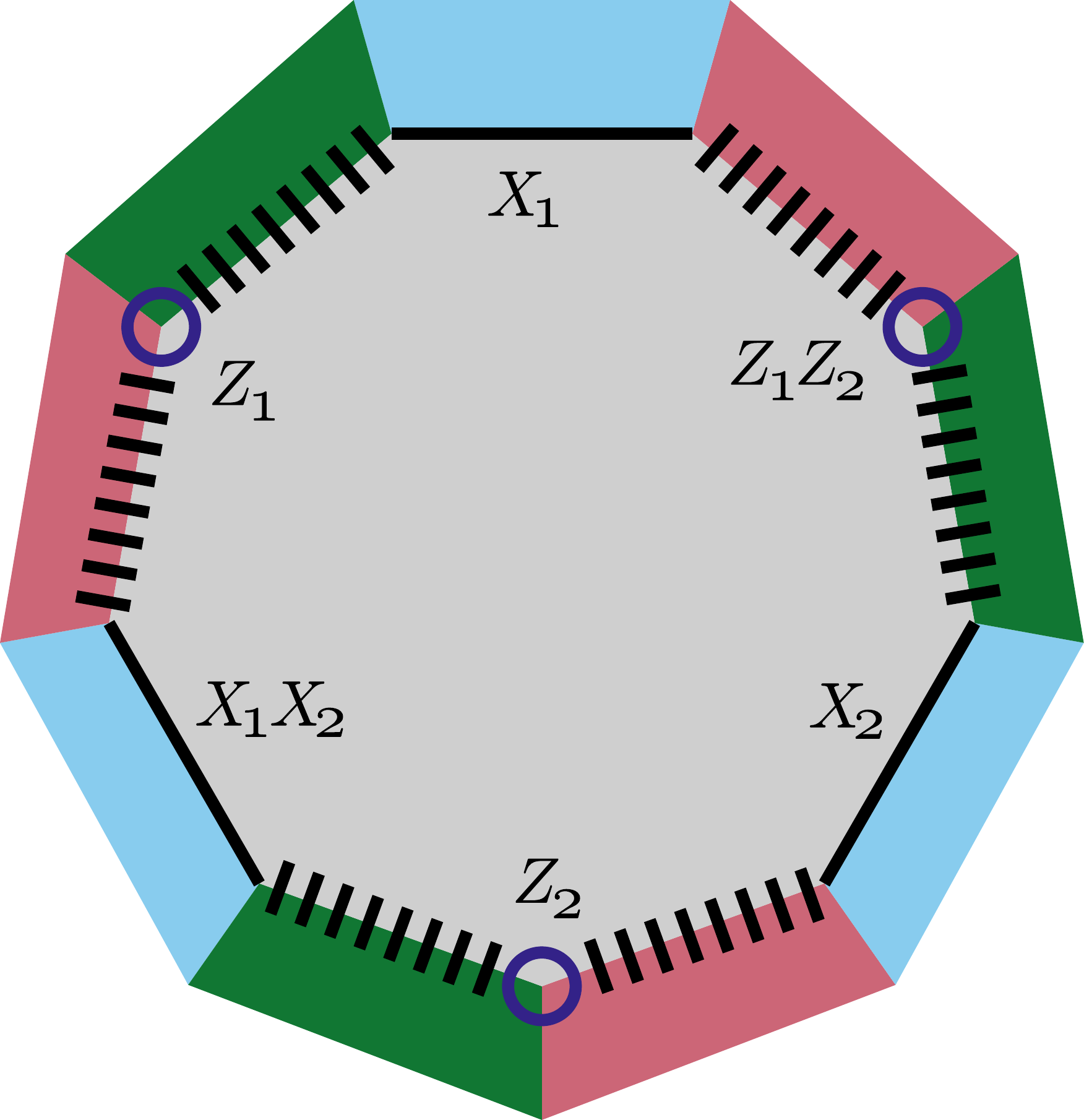}\\\(\qquad\qquad\qquad\qquad\Qcircuit @C=1em @R=.7em {
			& \gate{H} & \ctrl{1} & \qw \\
			& \gate{H} & \targ & \qw
		}\qquad\qquad\qquad\qquad\downarrow\)\\
		\includegraphics[width=.2\linewidth]{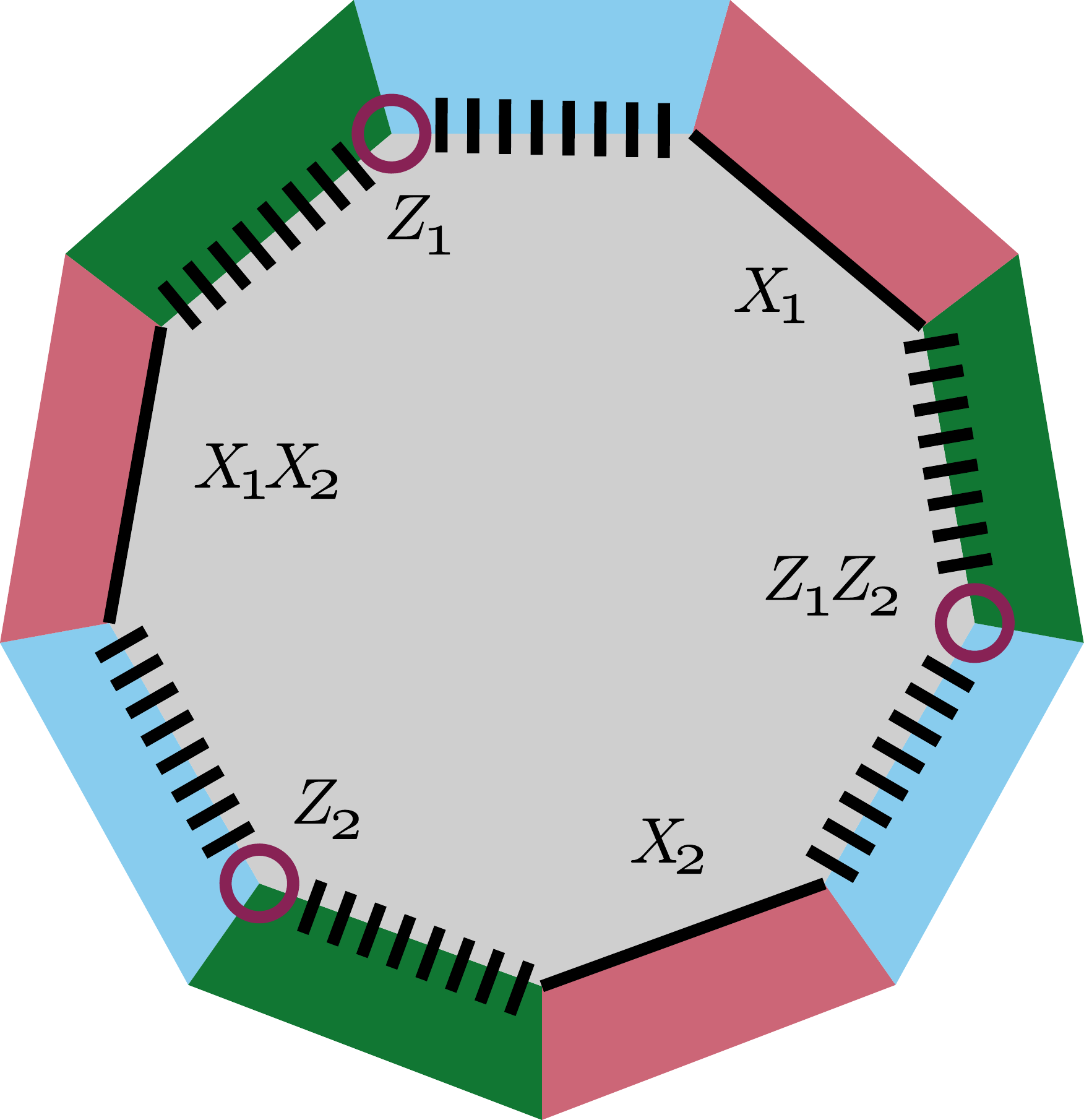}\hfil\raisebox{1.5cm}{\(\leftarrow\)}\hfil
		\includegraphics[width=.2\linewidth]{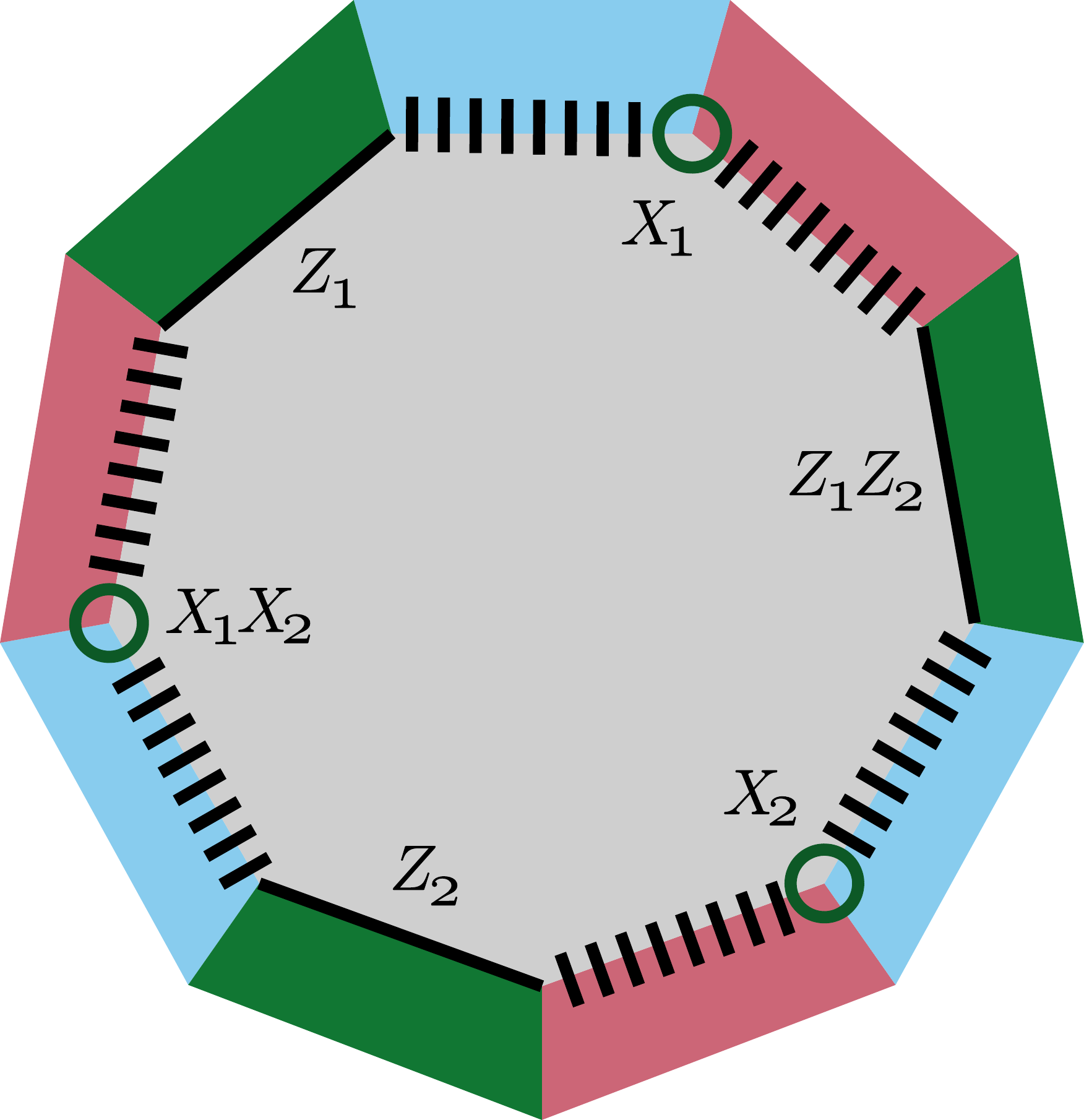}
		\caption{The dance of logical operators in a two-qubit planar Floquet code.
						We fix the basis of logical operators at the first step (R) and track them along with the measurement schedule.
						After three step the logical qubits have undergone a CNOT and two Hadamard gates. The logical circuit is shown in the center of the figure.
						}
		\label{fig:laride}
	\end{figure}

	We can now see how to perform these operations starting from single-qubit planar Floquet codes.
	The idea is to use code deformation in the form of plain surgery \cite{vuillot_code_2019, vuillot_fault-tolerant_2020} to merge two patches into a single one with the correct boundary configuration.
	This is schematized in Figure~\ref{fig:plainsurgery}.
	We put next to one another two hexagons with one rotated \(60^\circ\).
	In Figure~\ref{fig:plainsurgery} this makes a red boundary facing a green boundary.
	In order to merge them at the blue measurement step, one measures blue checks across the facing red and green boundaries instead of along them.
	The red and green checks all remain the same otherwise, except for on check near the green and red corners which disappear in the merge.
	We present the microscopic details of the procedure for two patches of minimum distance 5 in Figure~\ref{fig:CNOT}.
	
	\begin{figure}[ht]
		\centering
		\includegraphics[width=.3\linewidth]{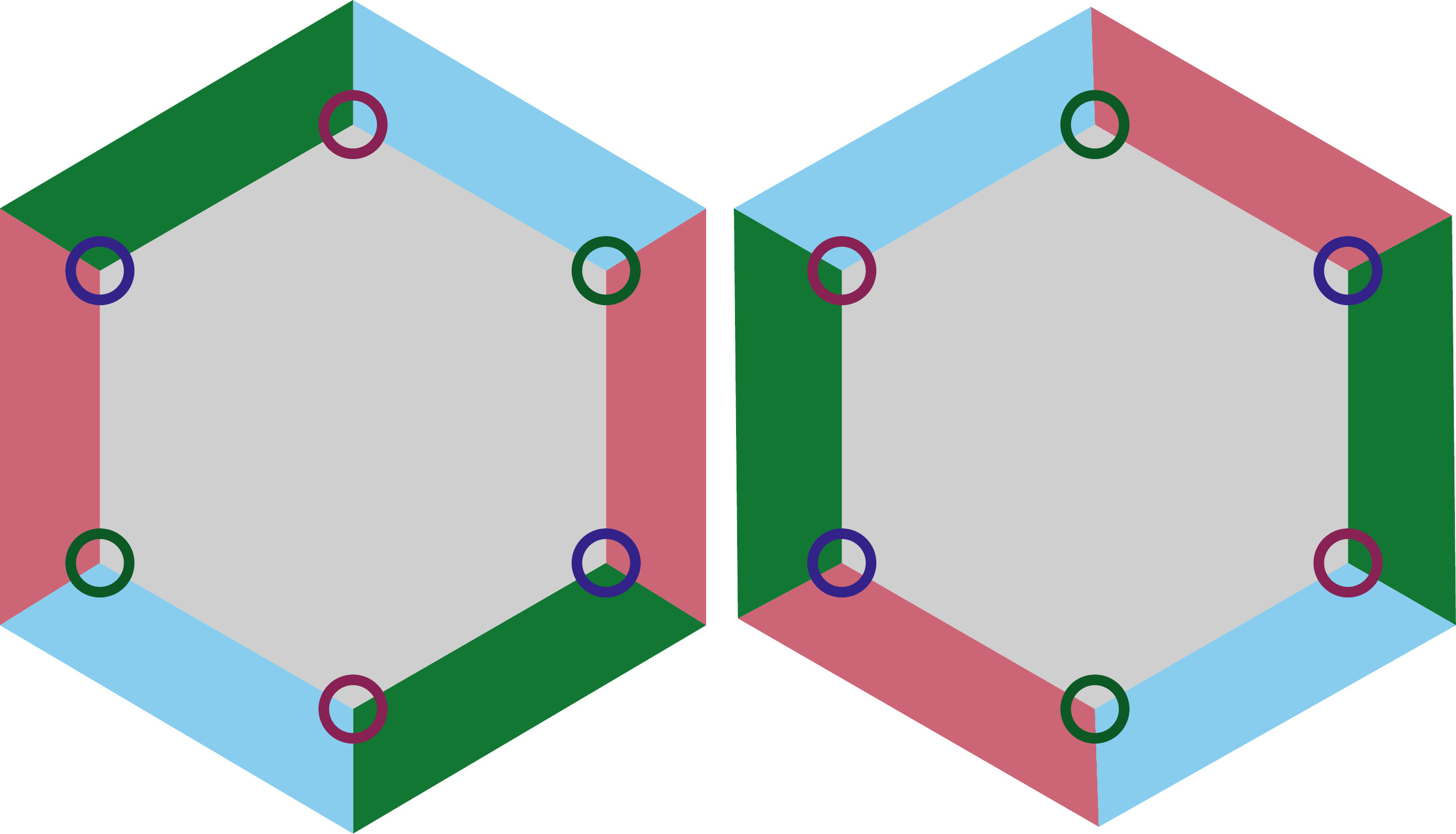}\hfil\raisebox{1.5cm}{\(\rightarrow\)}\hfil\includegraphics[width=.3\linewidth]{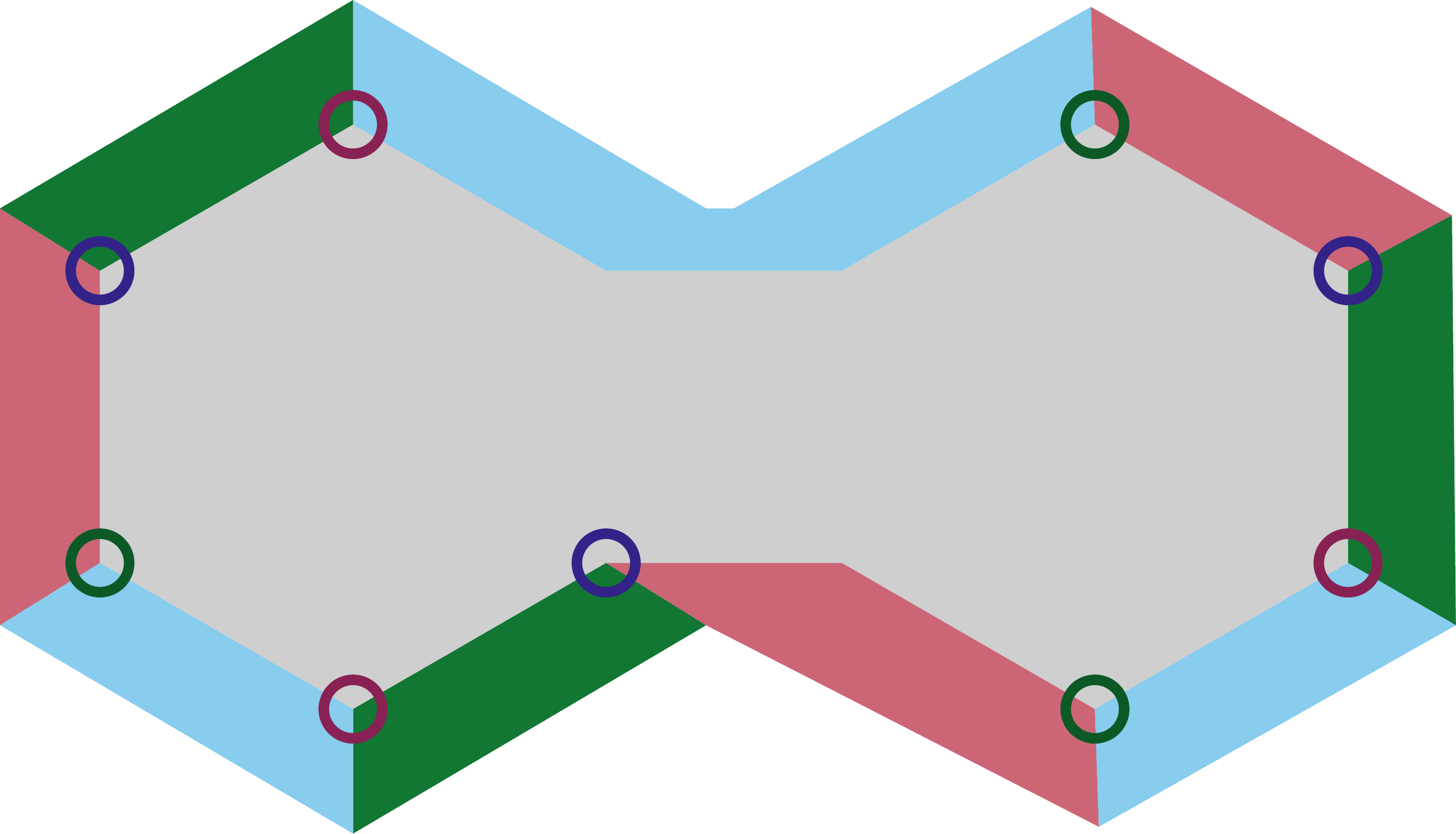}
		\caption{A configuration with two hexagonal planar Floquet codes that can be merged into a single nonagon.
					The logical information is kept untouched during the merging.
					The microscopic details of such plain merge are shown in Figure~\ref{fig:CNOT}}
		\label{fig:plainsurgery}
	\end{figure}

	One can check that the logical information of the two qubits is preserved by the merging and splitting procedures.
	For this it is enough to see that the logical operators hosted by the boundaries away from the merged boundaries are preserved and evolve as they would do without the merging procedure.
	When merging the patches by measuring blue (\(Z\)) gauge checks across boundaries, it creates a lot of random \(\pm\) signs on the new red (\(X\)) and green (\(Y\)) stabilizers extended across boundaries.
	This is equivalent to a high density of \(Z\) errors along the middle of the merged patch.
	The fault-tolerance of the procedure is then guaranteed by the fact that logical operators stringing \(Z\) operators, if they can terminate at the middle bottom of the merged patch, they have to reach to either right or left sides but cannot terminate at the top middle of the merged patch.
	Only strings of \(X\) or \(Y\) operators from top to bottom in the middle of the patch correspond to logical operators, see Figure~\ref{fig:FTmerge}.
	
	\begin{figure}[hb]
		\centering
		\includegraphics[width=.3\linewidth]{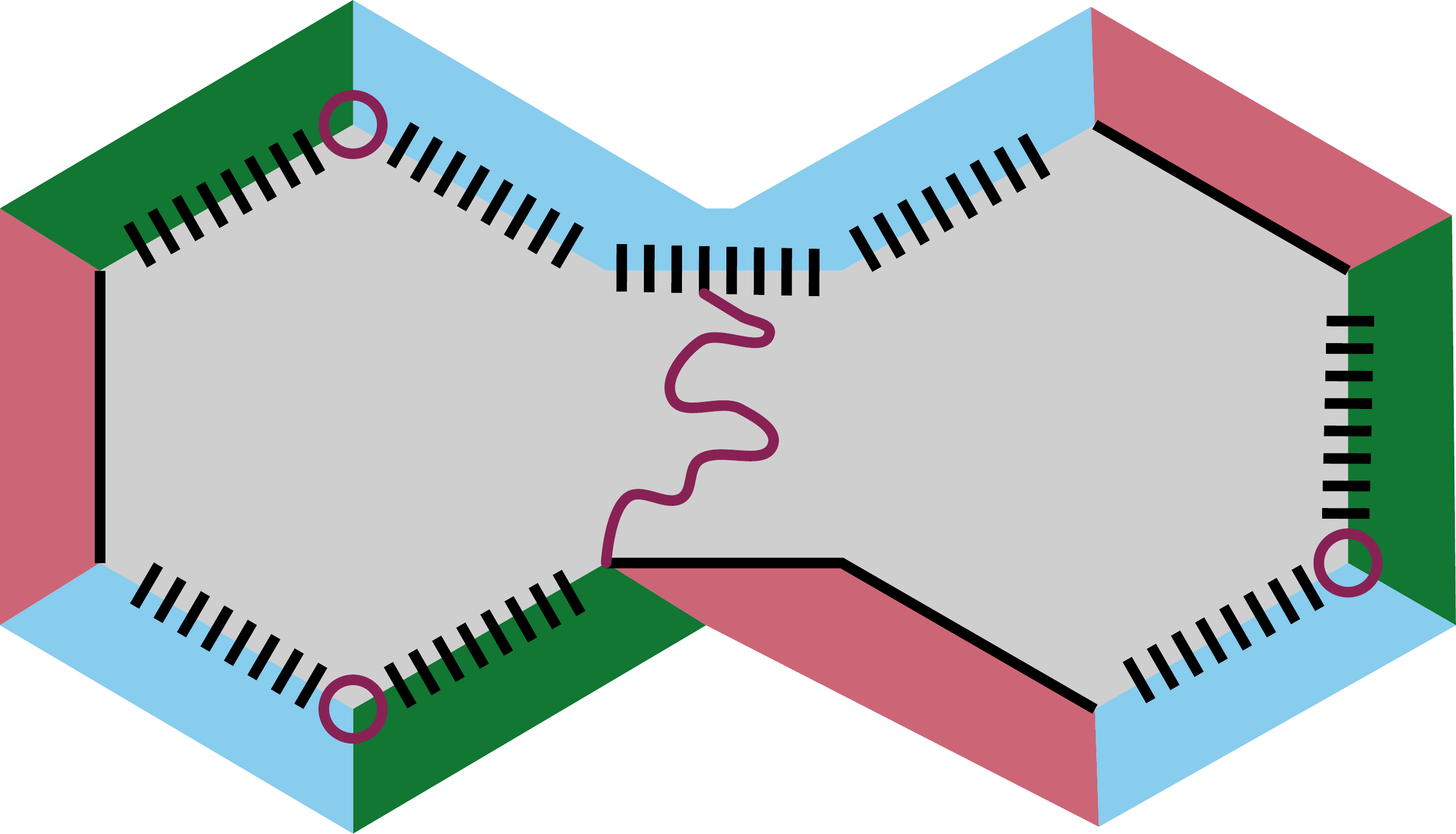}\hfil
		\includegraphics[width=.3\linewidth]{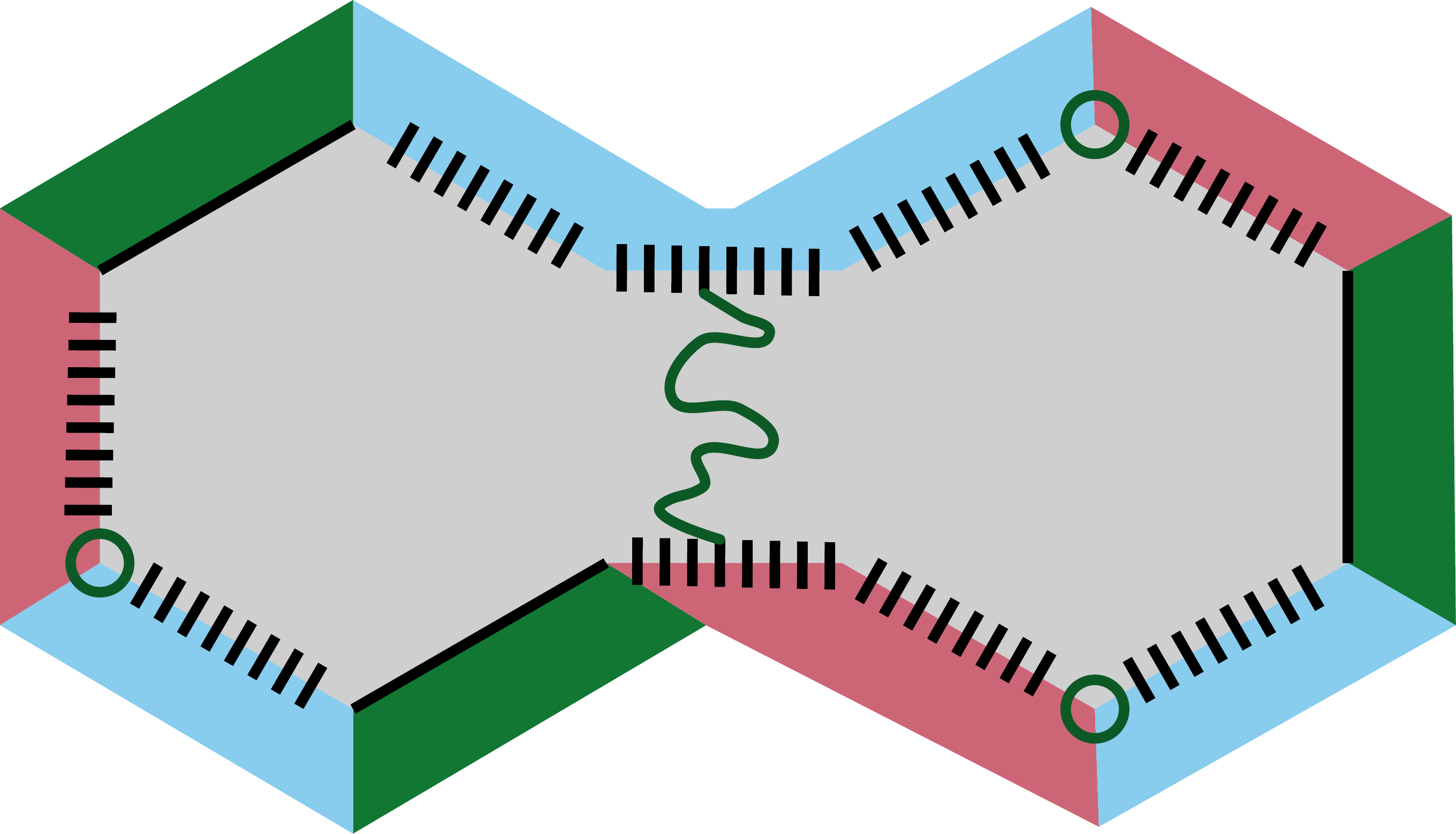}\hfil
		\includegraphics[width=.3\linewidth]{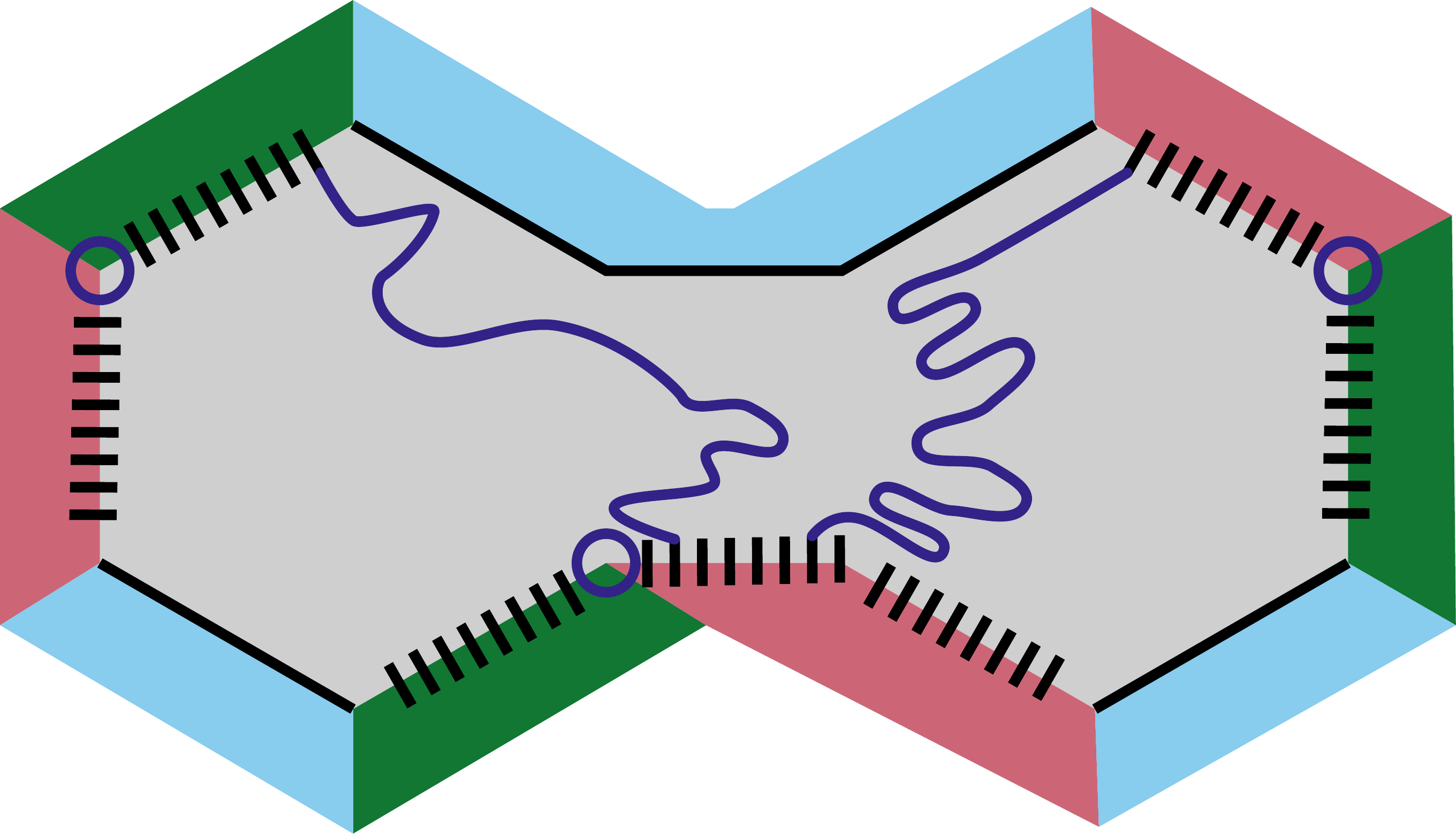}
		\caption{Two planar Floquet codes merged in a single patch.
				We show the different logical operators that can attach to the bottom middle of the patch.
				When red \(X\) checks are measured, a string of \(X\) operators down the middle is a logical operator. 
				When green \(Y\) checks are measured a string of \(Y\) operators down the middle is a logical operator.
				Strings of \(Z\) operators can eventually attach to the bottom middle but then has to terminate on the right or left side of the merged patch.
				So a high density of \(Z\) errors down the middle will not ruin the protection of the logical information.}
		\label{fig:FTmerge}
	\end{figure}

	\begin{figure}[p]
		\centering
		\includegraphics[width=.45\linewidth]{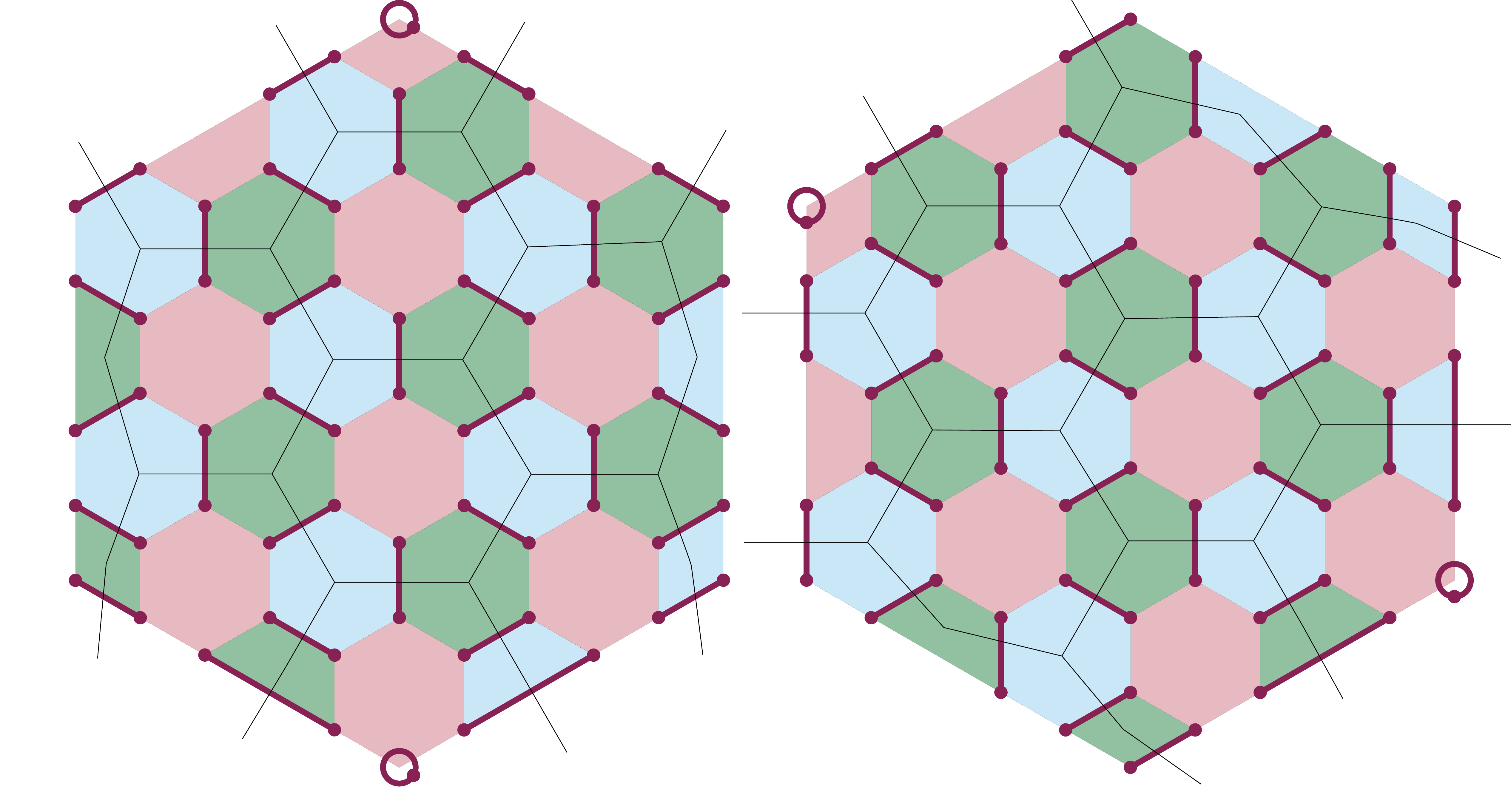}\\\(\downarrow\)\\
		\includegraphics[width=.45\linewidth]{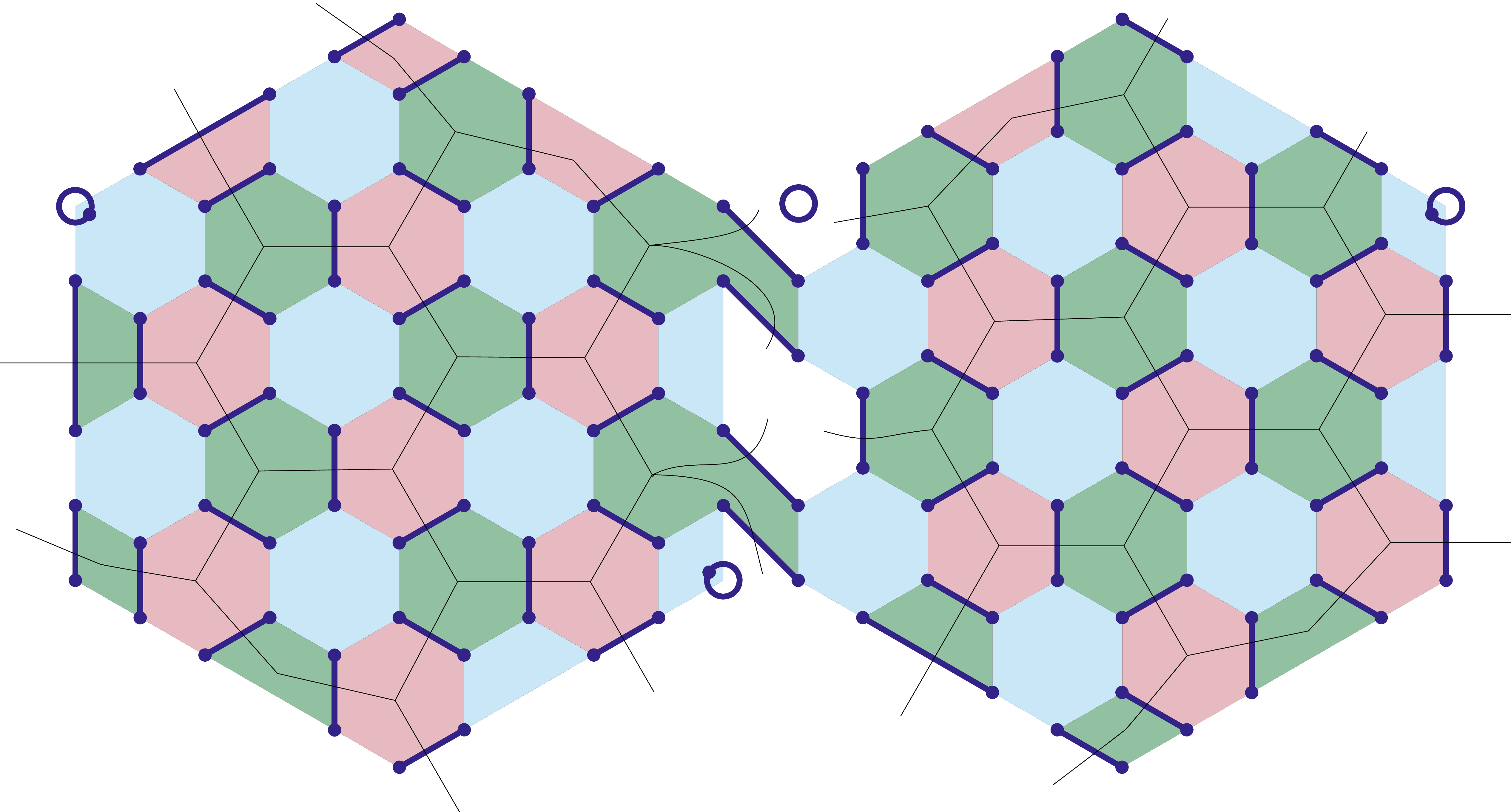}\\\(\searrow\)\\
		\includegraphics[width=.45\linewidth]{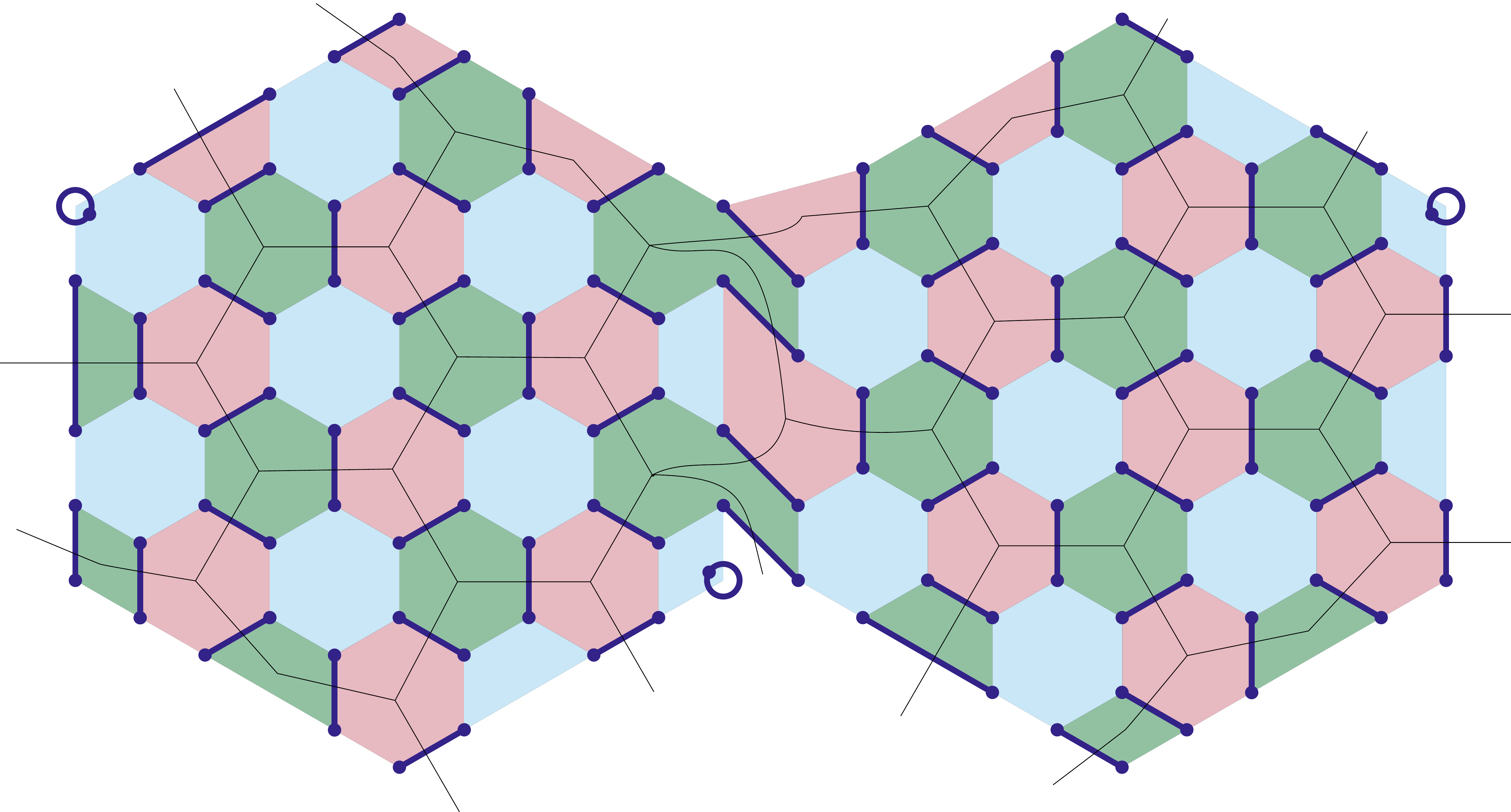}\hfill\raisebox{1.8cm}{\(\rightarrow\)}\hfill
		\includegraphics[width=.45\linewidth]{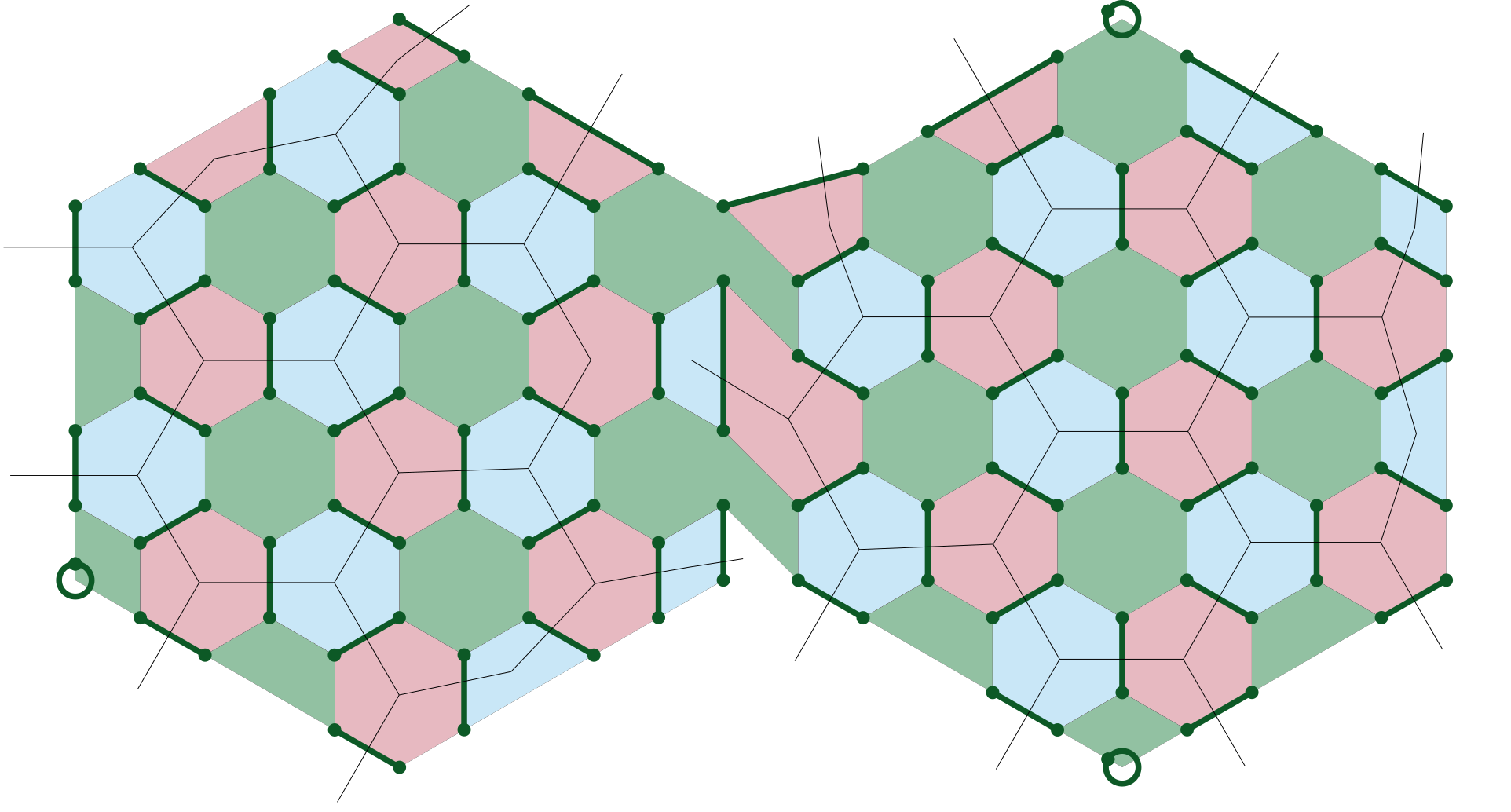}\\\(\nwarrow\qquad\qquad\swarrow\)\\[1em]
		\includegraphics[width=.45\linewidth]{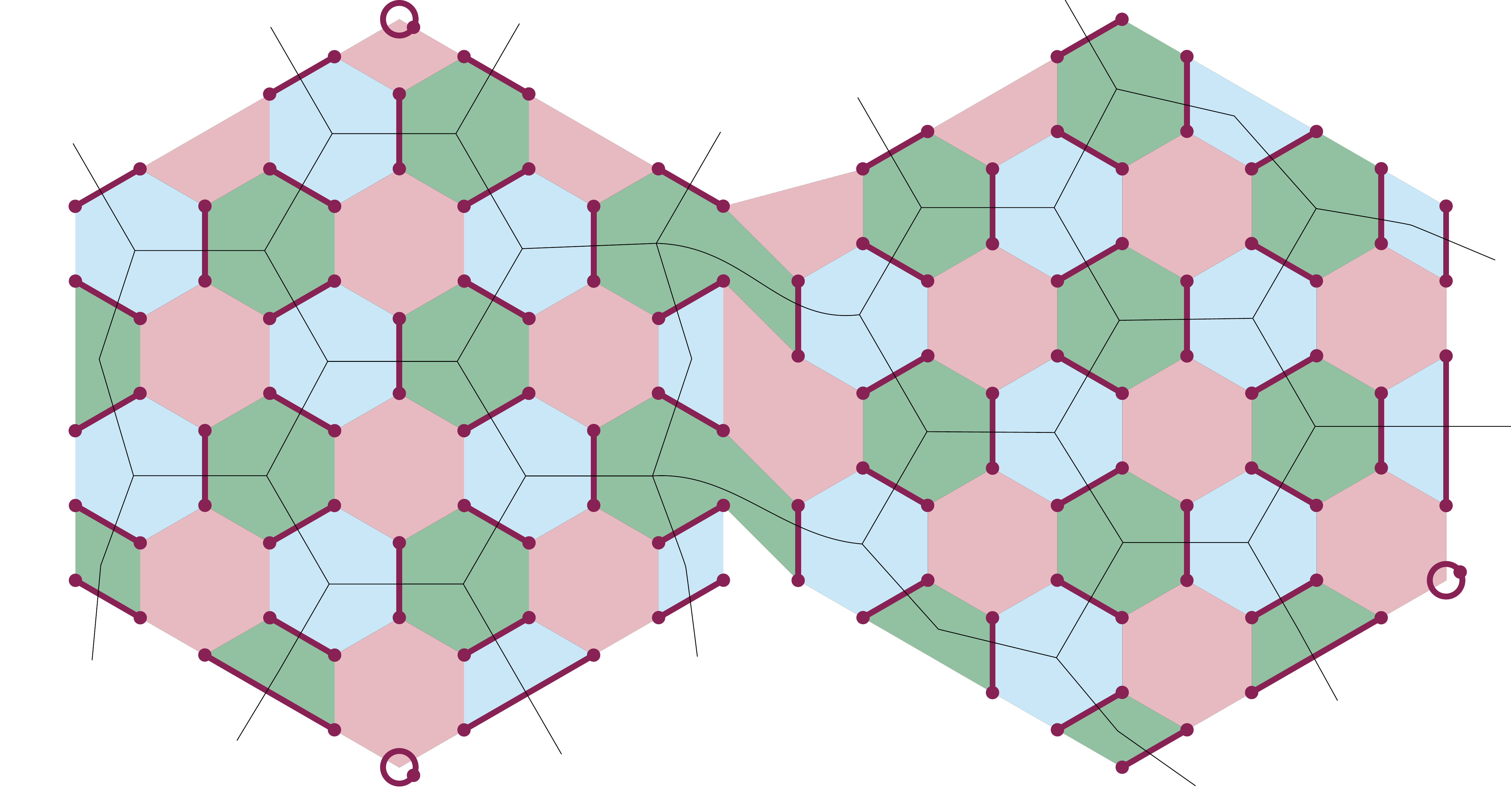}\\\(\downarrow\)\\
		\includegraphics[width=.45\linewidth]{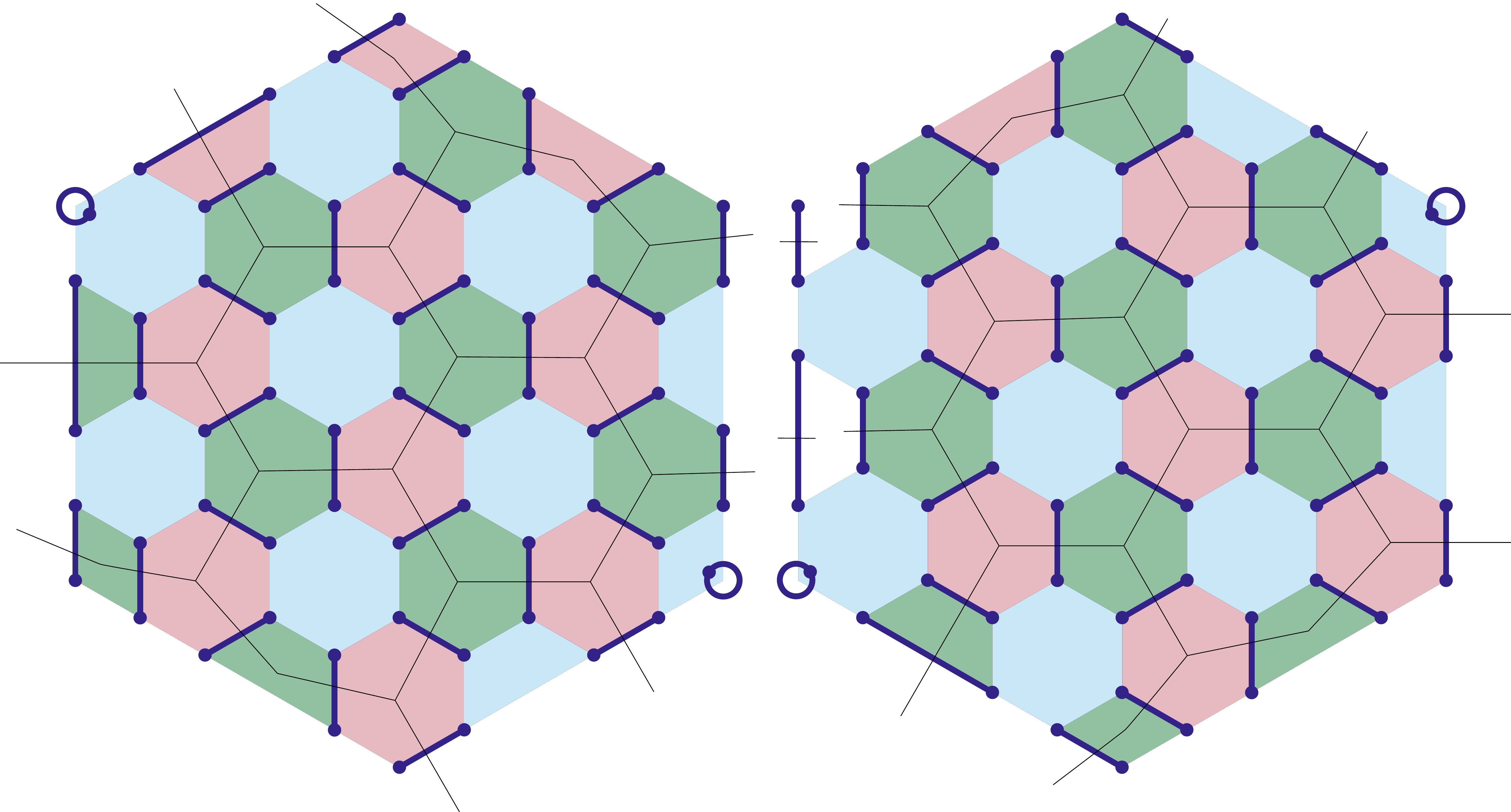}
		\caption{A schedule of measurements for realizing a CNOT.
		Arrows point toward the next set of check measurements.
		Note that the blue corner qubit of the right patch is measured out and removed from the patch at the second step.
		It is then reintroduced when splitting back the two patches.
		The loop in the middle is executed a number of times according to the desired operations on the logical qubits.
		Note also that at the merging step, which is a blue step, it helps to first measure the original blue checks and only then the new ones.
		This allows to measure for one last time the original green stabilizers before deforming them to merge the patches.
		The same thing can be done at the splitting blue step.}
		\label{fig:CNOT}
	\end{figure}
	
	\section{Conclusion}
	Building on the work of Hastings and Haah \cite{hastings_dynamically_2021}, we have defined Floquet codes on any 2D color code lattice \cite{bombin_topological_2006}.
	We have shown how to introduce some boundaries to the system taking the form of odd length colored boundaries of the color code \cite{kesselring_boundaries_2018}.
	The logical information in such planar Floquet codes gradually rotates geometrically and logically undergoes a Hadamard gate every three steps.
	For patches hosting more than one logical qubit, entangling operation are also regularly performed on the logical space.
	The rotation of the logical operators makes it so the overall protocol have constant-sized logical errors and is therefore not useful for fault-tolerant computation.
	We leave open the question of finding a way of keeping this rotating dynamics but also preserving the distance of the code.
	
	For set-ups allowing more flexibility in the connectivity, one could also use, for instance, tessellations for 2D hyperbolic color codes \cite{delfosse_tradeoffs_2013}.
	This would yield a constant rate Floquet codes with logarithmic distance.
	The decoding problem for these codes would be that of hyperbolic surface codes which can prove advantageous in terms of resources \cite{breuckmann_hyperbolic_2017}.
	Besides it is very likely that the construction generalizes to quantum pin codes \cite{vuillot_quantum_2019} which have much of the structure of color codes.
	
	In subsequent work \cite{haah_boundaries_2021}, a fault-tolerant way to introduce boundaries is presented.This could prove to be a promising route towards fault-tolerant quantum computation.
	Investigating their performance numerically \cite{gidney_fault-tolerant_2021} and comparing different geometries is the natural next step to assess if they can be more advantageous than surface code architectures in practice.

	\section*{Acknowledgments}
	C.V. would like to thank B.M. Terhal for many useful comments on a previous version of this manuscript.
	C.V. would like to thank J. Haah for discussions about the distance of the code leading to the realization that there are constant-sized logical operators.
	
	\bibliographystyle{unsrtnat}
	\bibliography{Floquet_codes_arxiv_v2}

\end{document}